\documentclass[aps,pre,twocolumn,footinbib,longbibliography,showpacs,groupedaddress,floatfix,superscriptaddress]{revtex4-2}%

\usepackage[dvipsnames]{xcolor}
\definecolor{linkcolor}{rgb}{0.3,0.3,1.0} %hyperlink
\usepackage[pdftex,colorlinks=true, linkcolor= linkcolor, citecolor= linkcolor, urlcolor= linkcolor, hyperindex=true,hyperfigures=true]{hyperref} %hyperlink

\usepackage{amssymb,amsmath,amsfonts}
\usepackage{graphicx}
\usepackage{float}
\usepackage{bm}
\usepackage{appendix}
\usepackage{verbatim}
\newcommand{\mean}[1]{\left\langle {#1} \right\rangle}

\newcommand{\traffic}{\mathcal{T}}
\newcommand{\inflow}{\mathcal{G}}
\newcommand{\xx}{{\bm{x}}}

\newcommand{\score}{{\bm{s}}}

\newcommand{\velo}{{\bm{v}}}
\newcommand{\JJ}{{\bm{J}}}
\newcommand{\FF}{\bm{F}}
\newcommand{\drift}{\bm{a}}
\newcommand{\xii}{\bm{\xi}}
\newcommand{\Tr}{\mathrm{Tr}}

\newcommand{\MA}{{\mathsf{A}}}

\newcommand{\MC}{{\mathsf{C}}}
\newcommand{\MD}{{\mathsf{D}}}

\newcommand{\MM}{{\mathsf{M}}}

\newcommand{\subs}{\Omega}% observed subset of degrees of freedom
\newcommand{\hidd}{H}% complementary hidden subset

\begin{document}

\title{CYNAR: a trajectory-based estimator of entropy production}

\author{Marco Baiesi}
\affiliation{Department of Physics and Astronomy, University of Padova,
Padova, Italy}
\affiliation{INFN, Sezione di Padova, Padova, Italy}
\affiliation{Max Planck Institute for the Physics of Complex Systems, Dresden, Germany}

\author{Ivan Di Terlizzi}
\affiliation{Max Planck Institute for the Physics of Complex Systems, Dresden, Germany}
\affiliation{Ludwig-Maximilians-Universit\"at M\"unchen, Arnold-Sommerfeld-Center for Theoretical Physics, M\"unchen, Germany}

\begin{abstract}
Inferring the entropy production rate $\sigma$ from recorded trajectories is a key challenge in stochastic thermodynamics. In experiments, the forces governing the dynamics are usually unknown and often only some of the system's degrees of freedom can be observed. Building on a recently introduced method [I.~Di Terlizzi, Phys.\ Rev.\ Lett.\ {\bf 135}, 237101 (2025)], we develop and validate CYNAR (computation yields nonequilibrium analysis and reconstruction), a pipeline that estimates $\sigma$ from trajectories. It exploits a kinetic decomposition of $\sigma$ into a \emph{traffic} term $\traffic$, inferred from the short-time curvature of correlation functions, and an \emph{inflow rate} $\inflow$, computed from the steady-state score, i.e., the gradient of the log-density. We show that the same decomposition, evaluated on any subset of observed coordinates, always bounds $\sigma$ from below, for any constant diffusion tensor and without knowledge of the hidden degrees of freedom. We then compare CYNAR with a more standard approach that estimates $\sigma$ from steady-state probability currents, on four systems of increasing complexity. CYNAR proves to be more reliable, with no detectable bias close to equilibrium, weak dependence on spatial discretization, and marked robustness to measurement noise. It is also readily applicable to high-dimensional systems, where the traffic alone provides a lower bound close to $\sigma$, and remains informative under partial observation, including cases where the flux-based estimate vanishes identically. CYNAR is provided as an open-source Python package.
\end{abstract}

\maketitle

\section{Introduction}

Living systems operate out of equilibrium and dissipate energy to move, to grow, and to process information. This dissipation leaves a signature in their fluctuations, which are no longer time-reversal symmetric. In a non-equilibrium steady state, the entropy production rate $\sigma$ quantifies both the heat dissipated into the environment and the breaking of time-reversal symmetry~\cite{maes03_1,seifert2012stochastic,peliti2021book,shiraishi2023book,GalCohenFluc,fluc1,kurchan1998fluctuation}. Measuring $\sigma$ is therefore a central problem in soft and biological matter. Indeed, broken detailed balance has been detected at scales ranging from single cells to the brain~\cite{martin2001comparison,battle2016broken,gladrow2016broken,turlier2016equilibrium,gnesotto2018broken,lynn2021broken,roldan2021quantifying,diterlizzi2024variance,manikandan2024estimate,edgar_perspective}, and is a defining feature of active matter~\cite{nardini2017entropy,dabelow2019irreversibility,loos2020irreversibility,fodor2022irreversibility,obyrne2022time,ro2022model}. Accordingly, many approaches have been proposed to quantify $\sigma$ from data, each exploiting different features of steady states~\cite{roldan2010estimating,seifert2019stochastic,lucente2025conceptual,seifert2026universal,ghosal2026identification}.

When the forces, the trajectory, and the diffusion constants of a system are known, stochastic energetics gives $\sigma$ through an established formula (see Eq.~\eqref{sigma-force} below)~\cite{sekimoto1998langevin,sek10,seifert2012stochastic,peliti2021book}. In experiments, however, the forces are usually unknown and only trajectories are recorded~\cite{turlier2016equilibrium,battle2016broken,manikandan2024estimate}. The Harada-Sasa relation circumvents this problem with additional perturbation experiments~\cite{har05,har06,toyabe2010nonequilibrium,fodor2016nonequilibrium,wang2018inferring}, while stochastic force inference first reconstructs the forces from the data~\cite{frishman2020learning,bruckner2020inferring,hem2025learning,gerardos2025principled}. This paper addresses the problem of estimating $\sigma$ directly from trajectories, assuming only a general overdamped diffusive dynamics.

The time irreversibility of trajectories can be used to infer $\sigma$ directly~\cite{andrieux2007entropy,roldan2010estimating,roldan2012entropy,li2019quantifying,martinez2019inferring,kim2020learning,ro2022model,kwon2024alpha,lanza2026,dallenogare2026}, but reliable estimates require trajectory lengths that grow quickly with $\sigma$. Thermodynamic uncertainty relations instead bound $\sigma$ from the fluctuations of observed currents~\cite{bar15,gin16,pietzonka2017finite,macieszczak2018unified,dit19,hasegawa2019fluctuation,hasegawa2019uncertainty,falasco2020unifying,horowitz2020thermodynamic,di2020thermodynamic,shiraishi2021optimal,van2022unified,pietzonka2024thermodynamic,puglisi2025brief} (a related, non-stationary bound follows from time-dependent moments~\cite{singh2024inferring}). The bound becomes tighter for suitably chosen currents~\cite{busiello2019hyperaccurate,van2020entropy,dieball2023direct}, which can be found with machine learning~\cite{manikandan2020inferring,otsubo2020estimating,manikandan2021quantitative,manikandan2024estimate,otsubo2022estimating,aguilera2026inferring,das2026localising,manikandan2026free}. Another strategy recasts the estimation of local thermodynamic functions as a learning problem on trajectory currents~\cite{lyu2024learning,lyu2026learning}.

All these estimates become harder under partial observation. Discrete systems may reveal only coarse-grained states or a subset of their transitions~\cite{seifert2026universal,bo2014entropy,bisker2017hierarchical,busiello2019entropy,teza2020exact,skinner2021estimating,skinner2021improved,harunari2022learn,van2022thermodynamic,ghosal2023entropy,blom2024milestoning,baiesi2024effective,fritz2025entropy,teza2025coarse,maier2026compensating}. Continuous processes may be coarse-grained in space~\cite{gingrich2017inferring,dieball2022mathematical,dieball2022coarse,ghosal2022inferring,fritz2026waiting,lanza2026} or have hidden degrees of freedom~\cite{gnesotto2020learning,roldan2021quantifying,lucente2022inference,busiello2024unraveling}. For a hair-bundle model closely related to the one we use below, hidden variables substantially degrade estimates based on time irreversibility~\cite{roldan2021quantifying}. 
When only a single coordinate is observable, bounds based on its noisy oscillations offer a complementary route, relating the entropy produced per period to the number of coherent oscillations~\cite{oberreiter2022universal,santolin2025dissipation,nagayama2025duality,kolchinsky2026elementary,gu2026dissipation,nagayama2026oscillatory}.

%Bounds based on the coherence of noisy oscillations offer a complementary route from a single observable coordinate, provided its autocorrelation displays visible oscillations. These bounds involve the number of coherent oscillations, namely the ratio between their decay time and period~\cite{oberreiter2022universal}. The entropy produced per period is bounded from below by $4\pi^2$ times this number for weak-noise limit cycles~\cite{santolin2025dissipation,nagayama2025duality}, for Gaussian phase-current fluctuations~\cite{kolchinsky2026elementary}, and, with a mode-dependent prefactor, for Markov jump processes~\cite{gu2026dissipation}, while linear dynamics obeys a twofold tighter bound~\cite{nagayama2026oscillatory}.

For Langevin systems, equilibrium corresponds to a vanishing irreversible component of the steady-state probability current, and in overdamped dynamics, where the whole current is irreversible, nonzero currents thus signal broken detailed balance~\cite{battle2016broken,gladrow2016broken,gnesotto2018broken}. These currents are also directly connected to $\sigma$. Dividing the current at each point in state space by the local probability density gives the probability velocity, that is, the mean velocity with which the system moves through that point. The steady-state average of the squared probability velocity, weighted by the inverse diffusion matrix, equals $\sigma$ (see Eq.~\eqref{sigmav2} below)~\cite{tome2010entropy,spinney2012entropy,gnesotto2018broken,gonzalez2019experimental}. To our knowledge, this approach, which we refer to as \emph{the $v^2$ method} throughout this paper, has mostly been used as a qualitative detector of nonequilibrium~\cite{battle2016broken,gladrow2016broken}, and seldom as a quantifier of entropy production.

A recently proposed method~\cite{diterlizzi2025force} takes a different route and is based on a decomposition of the entropy production rate, $\sigma=4\traffic+\inflow$, into a traffic $\traffic$ and an inflow rate $\inflow$~\cite{maes2008steady,MAES20201,baiesi2015inflow}. Both terms can be evaluated from position trajectories. The traffic can be estimated from the short-time behavior of correlation functions and can serve, for example, as an accessible estimator of dissipation in a partially observed chemical-sensing model~\cite{nicoletti2026balancing}. The inflow rate requires the steady-state score~\cite{hyvarinen2005estimation}, i.e., the gradient of the log-probability density, which is typically estimated with histograms on a spatial grid. This approach~\cite{diterlizzi2025force}, which we extend and make accessible through the ``computation yields nonequilibrium analysis and reconstruction'' (CYNAR) pipeline, is the main subject of this paper. We benchmark it against the $v^2$ method on four qualitatively different systems.

We find several appealing properties of CYNAR, which distinguish it from the $v^2$ method. i)~CYNAR is a difference of two terms and is not constrained to be non-negative, so its estimate can fluctuate on both sides of zero, which matters near equilibrium, where $\sigma\gtrsim 0$. The $v^2$ method, being a quadratic form, is instead non-negative and is therefore necessarily biased upward in this regime. ii)~Both the inflow rate and the $v^2$ method require partitioning the state space into cells of fixed size, which we use to build histograms of the density and of the local velocity. The CYNAR estimate of $\sigma$ is stable and varies slowly with the cell size, while the $v^2$ method depends sensitively on it. iii)~In high-dimensional systems, where a spatial grid becomes infeasible, the traffic term alone still provides a lower bound on $\sigma$, which we find to be close to $\sigma$ far from equilibrium. Hence, a usable estimate exists even when the inflow-rate contribution cannot be computed in practice. iv)~Because the traffic fit may use correlation lags excluding very short times, it is markedly more robust than the $v^2$ method to measurement noise with little temporal correlation. v)~Under partial observation of only some degrees of freedom, the CYNAR estimate computed from the observed coordinates remains a lower bound on $\sigma$ and is potentially informative. In contrast, the $v^2$ method's flux-based estimate is less effective and even vanishes when only one coordinate is observed.

The following section introduces the model assumptions and the kinetic inference of $\sigma$, presents an improved lower bound for partially observed systems, recalls the $v^2$ method, and relates the kinetic and velocity-based approaches to the standard drift-based formulas, before describing how the traffic, the inflow rate, and the local mean velocity are computed in practice. Then, Sec.~\ref{sec:examples} shows our examples, followed by a discussion in Sec.~\ref{sec:discussion} and conclusions in Sec.~\ref{sec:conclusions}. The derivation of the bound is given in Appendix~\ref{app:bound}, and the corresponding result for the $v^2$ method in Appendix~\ref{app:v2}.

\section{Model and methods}

\subsection{Multidimensional overdamped diffusion}
\label{sec:model}

We study the steady-state statistics of a system with $N$ degrees of freedom, whose state $\xx=(x^1,\ldots,x^N)$ evolves according to the overdamped stochastic differential equation
\begin{equation}
\label{LE}
\dot\xx_t = \drift(\xx_t) + \sqrt{2\MD}\,\xii_t \, .
\end{equation}
The drift $\drift(\xx)$ may include non-conservative forces, and $\xii_t$ is a white noise with mean $\mean{\xi^i_{t}}=0$ and covariance $\mean{\xi^i_{t}\xi^j_{s}}=\delta^{ij}\delta(t-s)$. The diffusion tensor $\MD$ is a constant, positive-definite, symmetric $N\times N$ matrix, not diagonal in general. For a system at temperature $T$ with mobility tensor $\MM$ and forces $\FF$, one has $\MD = \MM T$ and $\drift = \MM \FF$.

The probability density $\rho(\xx,t)$ associated to this dynamics obeys the Fokker-Planck equation
\begin{equation}
\label{FP}
\partial_t \rho(\xx,t) = -\nabla\cdot\JJ(\xx,t) \, ,
\qquad
\JJ = \drift\,\rho - \MD\nabla\rho \, ,
\end{equation}
where $\JJ$ is the probability current. We assume that the system relaxes to a steady state with density $\rho(\xx)$, so that the correlation functions used below are well defined. In the steady state, $\partial_t\rho=0$ implies that the current is divergence-free, $\nabla\cdot\JJ=0$. Equilibrium corresponds to a vanishing current, $\JJ=0$, which is the condition of detailed balance. In a nonequilibrium steady state, instead, a nonzero current circulates in state space. Dividing it by the density gives the local mean velocity
\begin{equation}
\label{velocity}
\velo(\xx) = \frac{\JJ(\xx)}{\rho(\xx)} = \drift(\xx) - \MD\,\nabla\ln\rho(\xx) \, ,
\end{equation}
which vanishes everywhere in equilibrium.

Throughout the text, we refer to a trajectory $\{\xx\} = (\xx(dt),\xx(2\,dt),\ldots,\xx(n\,dt))$ as a sequence of $n$ data points, obtained by sampling the state $\xx_t$ at constant time step $dt$. This is the only information available to our estimators.

\subsection{Kinetic inference of entropy production}
\label{sec:overview}

Our starting point is a decomposition of the steady-state entropy production rate introduced by Maes and coworkers~\cite{maes2008steady,MAES20201},
\begin{align}
\label{sigma-Maes}
\sigma = 4\traffic + \inflow \, ,
\end{align}
in which both terms are averages of time-symmetric quantities. The traffic $\traffic$, also called dynamical activity or frenesy~\cite{MAES20201}, is the time-symmetric part of the path probability, taken relative to a drift-free diffusion with the same noise~\cite{maes2008steady}.  The inflow rate $\inflow$~\cite{baiesi2015inflow}, instead, measures the mean contraction of the probability in state space due to the drift. We recall their expressions in terms of $\drift$ in Sec.~\ref{sec:drift}. Here, we follow Ref.~\cite{diterlizzi2025force}, which recasts both terms in forms that only require the trajectory $\{\xx\}$.

The first ingredient is the connected correlation matrix $\MC(t)$ between degrees of freedom, evaluated at $t=0,dt,2\,dt,\ldots$, with elements
\begin{align}
    C^{ij}(t) = \mean{x^i(t) x^{j}(0)} - \mean{x^i(t)}\mean{x^{j}(0)}\,,
\end{align}
where $\mean{\ldots}$ denotes the steady-state average.
The time derivative of its symmetrized version,
\begin{align}
  \MC_* = \frac 1 2 \left(\MC + \MC^T\right) ,
\end{align}
yields the diffusion tensor,
\begin{align}
\label{Dhat}
    \MD = -\dot\MC_*(0^+)\,.
\end{align}
The second time derivative $\ddot\MC_*(0^+)$ then gives the traffic,
\begin{align}
\label{traffic}
\traffic
&= -\frac 1 4 \sum_{i,j}(\MD^{-1})^{ij}\,\ddot C_*^{ij}(0^+)\nonumber\\
&= -\frac 1 4 \Tr\left[\MD^{-1}\ddot\MC_*(0^+)\right]\,,
\end{align}
a result that follows from the variance sum rule~\cite{diterlizzi2024variance,diterlizzi2024vsrmodels,diterlizzi2025force} (see Sec.~\ref{sec:drift}). Most importantly, $\MD$ and $\ddot\MC_*(0^+)$ can be estimated from a finite trajectory, as described in Sec.~\ref{ssec:traffic}, so that traffic estimation requires only the temporal correlations of the measured degrees of freedom.

The inflow rate $\inflow$, instead, depends on the spatial distribution $\rho(\xx)$ in the steady state, through the gradient of its logarithm,
\begin{align}
    \score(\xx) = \nabla \ln \rho(\xx) \, .
\end{align}
This score function plays a central role in many machine learning methods~\cite{hyvarinen2005estimation}, which in turn help estimate nonequilibrium response~\cite{klinger2025computing}. The inflow rate reads~\cite{frishman2020learning,diterlizzi2025force}
\begin{align}
\label{inflow}
    \inflow = \mean{\score \cdot\MD \score} = \sum_{i,j}\int d\xx \,\rho(\xx)\,s^i(\xx)D^{ij}s^j(\xx) \, ,
\end{align}
where, again, we use the estimate of $\MD$ from Eq.~\eqref{Dhat}.

Combining Eqs.~\eqref{traffic} and \eqref{inflow}, the entropy production rate in the steady state is
\begin{align}\label{sigma}
    \sigma &= 4\traffic + \inflow \nonumber\\
    & = -\Tr\left[\MD^{-1}\ddot\MC_*(0^+)\right] + \mean{\score \cdot \MD \score} \, .
\end{align}
Being a difference of two terms, the estimate~\eqref{sigma} is not constrained to be non-negative and can take negative values, which should then be compatible with the equilibrium value $\sigma=0$ within error bars. The estimator is therefore potentially unbiased near equilibrium. Indeed, at equilibrium, $4\traffic+\inflow=0$, and since the inflow rate~\eqref{inflow} is a weighted covariance of the score and hence positive, the traffic is negative, $\traffic=-\inflow/4<0$. By Eq.~\eqref{traffic}, this corresponds to a dominant positive curvature of the correlation functions at short times, consistent with the typical negative curvature of mean-square displacements~\cite{reichert2021tracer}. Far from equilibrium, instead, the traffic is often positive, reflecting significant negative curvatures, $\ddot C_*^{ij}(0^+)<0$.

\subsection{Improved bound from partially observed trajectories}
\label{sec:bound}

To address the practical challenge of estimating dissipation in a partially observed system, Ref.~\cite{diterlizzi2025force} proposed a lower bound based on a subset $\subs$ of observed degrees of freedom (improving a previous argument based on the variance sum rule~\cite{diterlizzi2024vsrmodels}). Here, we generalize and improve this bound (see Appendix~\ref{app:bound} for technical details). We define the traffic restricted to $\subs$ as
\begin{align}
\label{TS}
\traffic^\subs = -\frac 1 4 \sum_{i,j\in \subs}(\MD_\subs^{-1})^{ij}\,\ddot C_*^{ij}(0^+) \, ,
\end{align}
which requires only the correlations of the observed coordinates and the corresponding block $\MD_\subs$ of the diffusion tensor, obtained from the same correlations through Eq.~\eqref{Dhat}. Similarly, the inflow rate restricted to $\subs$,
\begin{align}
\label{GS}
\inflow^\subs = \mean{\tilde\score\cdot\MD_\subs\tilde\score} \, ,
\end{align}
is computed through the score $\tilde\score=\nabla_\subs\ln\rho_\subs$ of the marginal density $\rho_\subs(\xx_\subs)$ of the observed coordinates. As shown in Appendix~\ref{app:bound}, these two quantities bound from below the dissipation associated with the observed degrees of freedom,
\begin{align}
\label{sigmaS}
\sigma^\subs = \mean{\velo_\subs\cdot\MD_\subs^{-1}\velo_\subs} \, ,
\end{align}
where $\velo_\subs$ contains the observed components of the local mean velocity~\eqref{velocity} of the whole system, which depends on all degrees of freedom. Since $\sigma\ge\sigma^\subs$, we obtain
\begin{align}
\label{lowerTS}
\sigma \;\ge\; \sigma^\subs \;\ge\; \max\left(4\traffic^\subs+\inflow^\subs,\,0\right) \;\ge\; \max\left(4\traffic^\subs,\,0\right) ,
\end{align}
for any positive-definite $\MD$, including noise correlations between observed and hidden degrees of freedom. Unlike the two bounds on the right, $\sigma^\subs$ cannot be computed from the observed trajectory alone, since $\velo_\subs$ depends on the hidden coordinates. The first two inequalities become equalities when all degrees of freedom are observed, while the last follows from $\inflow^\subs\ge0$ and becomes particularly useful when only the traffic can be computed.

When $\MD_\subs$ is block diagonal with respect to a partition of $\subs$ into subsets $\subs_k$, i.e., when the noise does not correlate observed coordinates belonging to different subsets, the bound can be applied to each subset separately,
\begin{align}
\label{lowerT}
\sigma \;\ge\; \sum_{k}\max\left(4\traffic^{\subs_k}+\inflow^{\subs_k},\,0\right) ,
\end{align}
with $\traffic^{\subs_k}$ and $\inflow^{\subs_k}$ given by Eqs.~\eqref{TS} and \eqref{GS} restricted to $\subs_k$. Since $\MD_\subs$ is estimated from the observed correlations, this condition can be checked directly from the data. No condition is required on the noise correlations between observed and hidden degrees of freedom. Compared with Eq.~\eqref{lowerTS}, Eq.~\eqref{lowerT} only keeps the positive contributions of individual subsets, but each $\inflow^{\subs_k}$ is computed from a lower-dimensional marginal density, which carries less Fisher information. Neither bound is tighter in general, so the larger one applies. For a diagonal $\MD_\subs$, the subsets are reduced to single coordinates, with $\traffic^i = -\frac14 (D^{ii})^{-1}\ddot C_*^{ii}(0^+)$ and $\inflow^i = D^{ii}\mean{(\partial_i\ln\rho_i)^2}$, where $\rho_i$ is the marginal density of $x^i$. When the (positive) contribution of the inflow rate is not taken into account, one recovers the bound $\sigma\ge\sum_{i\in \subs}\max(4\traffic^i,0)$ of Ref.~\cite{diterlizzi2025force}, which involves only the traffic.

From a practical point of view, the only assumption needed to apply the results above to real data is that the whole system follows the overdamped dynamics~\eqref{LE} with a constant diffusion tensor and has a well-defined steady state. No knowledge is required about how many degrees of freedom are hidden or about their noise correlations with the observed coordinates, so that $\max(4\traffic^\subs+\inflow^\subs,0)$ bounds $\sigma$ from below for any set of recorded coordinates. The quality of the result then depends on how accurately $\traffic^\subs$ and $\inflow^\subs$ can be estimated from a finite trajectory, which poses different challenges for the two terms. In particular, the traffic requires the short-time curvature of noisy correlation functions, while the inflow rate requires the score of a density reconstructed on a spatial grid. The latter becomes increasingly difficult as the number of observed degrees of freedom grows, since the number of cells to populate increases exponentially with the dimension. In such cases, the bound $\max(4\traffic^\subs,0)$ based solely on the traffic remains a valuable option, as it only requires pairwise correlation functions. Moreover, far from equilibrium, the traffic typically carries most of the information on $\sigma$, since it grows with the strength of nonequilibrium driving, while the inflow rate is often comparatively low. We address these practical issues in Secs.~\ref{ssec:traffic}, \ref{sec:practice-inflow}, and \ref{sec:practice-v2}. Before that, Sec.~\ref{sec:v2method} introduces the $v^2$ method, which we use as a comparison baseline, and Sec.~\ref{sec:drift} relates the formulas above to the standard expressions based on the drift.

\subsection{Velocity-based method}
\label{sec:v2method}

An alternative route to detect and quantify broken detailed balance from trajectory data relies on the steady-state probability current~\cite{gnesotto2018broken}. In terms of the local mean velocity $\velo(\xx)$ of Eq.~\eqref{velocity}, the entropy production rate reads~\cite{tome2010entropy,spinney2012entropy,seifert2012stochastic}
\begin{align}
\label{sigmav2}
\sigma_{v^2} = \mean{\velo\cdot\MD^{-1}\velo} \, .
\end{align}
This expression is exact for the dynamics~\eqref{LE}. Like the kinetic inference of Sec.~\ref{sec:overview}, it does not require the drift, because $\velo$ can be estimated directly from the recorded trajectory, as described in Sec.~\ref{sec:practice-v2}. We refer to this estimator as the $v^2$ method. Unlike $4\traffic+\inflow$, however, Eq.~\eqref{sigmav2} is a quadratic form and is therefore non-negative for any estimate of $\velo$.

Under partial observation, the displacements of the observed coordinates only give access to the marginal current $\tilde\JJ_\subs=\int d\xx_\hidd\,\JJ_\subs$, i.e., to the marginal velocity $\tilde\velo=\tilde\JJ_\subs/\rho_\subs=\mean{\velo_\subs\,|\,\xx_\subs}$, which is the conditional average of the observed components of the full velocity field. The $v^2$ method applied to $\xx_\subs$ then yields
\begin{align}
\label{sigmav2S}
\sigma_{v^2}^\subs = \mean{\tilde\velo\cdot\MD_\subs^{-1}\tilde\velo} \;\le\; \sigma^\subs \;\le\; \sigma \, ,
\end{align}
where the first inequality follows from averaging the velocity over the hidden coordinates (see Appendix~\ref{app:v2}). Both the $v^2$ method and CYNAR thus bound from below the dissipation $\sigma^\subs$ of Eq.~\eqref{sigmaS}, but they lose information in different ways. The $v^2$ method discards the fluctuations of $\velo_\subs$ at fixed $\xx_\subs$, which can be large when the hidden coordinates drive the observed ones. CYNAR, instead, retains the traffic $\traffic^\subs$, which enters $\sigma^\subs$ without any loss, and only loses Fisher information through the marginal inflow rate. The corresponding gap cannot exceed $\inflow-\inflow^\subs$, the part of the inflow rate lost upon marginalization (see Eq.~\eqref{app:gap}), so that the CYNAR estimate is expected to be close to $\sigma^\subs$ far from equilibrium, where $\inflow\ll\sigma$. In the extreme case of a single observed coordinate, the marginal current vanishes identically with natural boundary conditions, so that $\sigma_{v^2}^\subs=0$, whereas $4\traffic^\subs+\inflow^\subs$ can remain positive.

\subsection{Relation to drift-based formulas}
\label{sec:drift}

When the drift $\drift$ is known, as in all simulated systems used below, $\sigma$ follows from the standard formula of stochastic energetics~\cite{sekimoto1998langevin,sek10,peliti2021book},
\begin{align}
\label{sigma-force}
\sigma &= \mean{(\MD^{-1}\drift)\circ\dot\xx}
\\
\label{sigma-force-discr}
& \simeq \frac{1}{n-1}\sum_{k=1}^{n-1} \left(\MD^{-1}\frac{\drift(\xx_k)+\drift(\xx_{k+1})}{2}\right) \cdot\frac{\xx_{k+1} - \xx_{k}}{dt} \, ,
\end{align}
namely a Stratonovich average (denoted by $\circ$) of the drift times the velocity $\dot\xx$. 

In the decomposition~\eqref{sigma-Maes} of Maes \emph{et al.}~\cite{maes2008steady}, the traffic and the inflow rate are expressed through the drift as
\begin{align}
\label{traffic-force}
\traffic &= \frac14\mean{\drift\cdot\MD^{-1}\drift} + \frac12\mean{\nabla\cdot\drift} \, ,\\
\label{inflow-force}
\inflow &= -\mean{\nabla\cdot\drift} \, .
\end{align}
The inflow rate~\eqref{inflow-force} is the mean contraction rate of the state space due to the drift~\cite{baiesi2015inflow}. Summing the two contributions as in Eq.~\eqref{sigma-Maes} gives
\begin{align}
\label{sigma-inst}
\sigma = \mean{\drift\cdot\MD^{-1}\drift} + \mean{\nabla\cdot\drift} \, ,
\end{align}
which recovers the alternative form of Eq.~\eqref{sigma-force} based on instantaneous averages, already provided by Sekimoto~\cite{sek10}. Compared with the Stratonovich average~\eqref{sigma-force}, it does not depend on the sampling rate, but it requires the gradients of the drift.

The formulae in Sec.~\ref{sec:overview} follow from these expressions. For the inflow rate, an integration by parts gives $-\mean{\nabla\cdot\drift}=\mean{\drift\cdot\score}$. Moreover, $\mean{\velo\cdot\score}=\int d\xx\,\JJ\cdot\nabla\ln\rho$ vanishes after a further integration by parts, since $\nabla\cdot\JJ=0$. Using $\velo=\drift-\MD\score$, we thus obtain $\inflow=\mean{\drift\cdot\score}=\mean{\score\cdot\MD\score}$, i.e., Eq.~\eqref{inflow}. For the traffic, the short-time curvature of the correlation functions reads
\begin{align}
\label{Cddot-drift}
\ddot C^{ij}(0^+) = -\mean{a^i a^j} + 2\mean{a^i\,(\MD\score)^j}
\end{align}
(see Appendix~\ref{app:bound}). Inserting its symmetrized version into Eq.~\eqref{traffic} and using $\mean{\drift\cdot\score}=-\mean{\nabla\cdot\drift}$ recovers Eq.~\eqref{traffic-force}. Combining the latter with Eq.~\eqref{sigma-inst} yields
\begin{equation}\label{sigma-VSR}
    \begin{split}
        \sigma  &= 2\traffic + \frac12\mean{\drift\cdot\MD^{-1}\drift} \\
        &= \frac12\Tr\left[\MD^{-1}\left(-\ddot\MC_*(0^+)+\mean{\drift\,\drift^T}\right)\right],
    \end{split}
\end{equation}
namely the formula derived from the variance sum rule in Refs.~\cite{diterlizzi2024variance,diterlizzi2024vsrmodels}. This relation first revealed that the short-time curvature of correlation functions is informative about dissipation, but its evaluation still requires the drift covariance. The same holds for Eqs.~\eqref{sigma-force}--\eqref{sigma-inst}, which cannot be applied when only positions are recorded. The kinetic inference implemented in CYNAR overcomes this limitation, since Eqs.~\eqref{traffic} and \eqref{inflow} only use the recorded positions, as does the $v^2$ method through Eq.~\eqref{sigmav2}. The drift-based formulas remain useful in our benchmarks, where the forces are known and provide reference values. For every nonlinear example below, we evaluate $\traffic$ and $\inflow$ from Eqs.~\eqref{traffic-force} and \eqref{inflow-force}, and $\sigma$ independently from Eq.~\eqref{sigma-force-discr}, as averages along one long simulated trajectory. The agreement between $\sigma$ and $4\traffic+\inflow$ from the same run then provides a self-consistency check of Eq.~\eqref{sigma-Maes}.

\subsection{Computing the traffic}
\label{ssec:traffic}

Let us focus on one of the $N\times N$ sequences $C^{ij}(t)$ for $t=k\,dt$ with integer $k\ge 0$, representing the temporal correlation between $x^{i}(t_0 + t)$ and $x^j(t_0)$ for arbitrary $t_0$. We can compute this correlation easily with a fast Fourier transform. Since we are interested in the first two time derivatives of $C^{ij}(t)$ for $t\to 0^+$, we need a method that yields the slope and curvature of the data. This method should not rely on a few values of $C^{ij}(t)$ for small $t$ because they would yield an approximation of the derivatives that is overly sensitive to potential high-frequency instrumental noise in the acquisition of experimental data. For this reason, it is convenient to fit the data with a function and get its derivatives analytically with the fitted parameters~\cite{diterlizzi2025force}.

A sensible function for fitting correlations should be a compromise between flexibility (for covering a variety of typical cases) and simplicity (to avoid overfitting). Typical cases include strong correlations that decay either exponentially or with initial negative curvature, and weak correlations that grow polynomially. We found that a linear superposition of $N_e$ functions
\begin{align}
\label{exppoly}
    B_\ell(t) = e^{-s_\ell t-r_\ell t^2}\left(b_\ell^{(0)}+b_\ell^{(1)}t+\ldots +b_\ell^{(N_g)} t^{N_g}\right)
\end{align}
works well in many cases, with either one ($\ell = N_e=1$) or at most two ($\ell\le N_e=2$) functions of this kind, and a polynomial of degree $N_g=2$.

Only lags with index $k\ge k_1$ enter the fit. This small offset (by default, $k_1=2$) excludes the earliest lags, where the estimate of short-time derivatives is most sensitive to measurement noise, an issue also encountered in related trajectory-inference schemes~\cite{frishman2020learning,ferretti2020building,bruckner2020inferring}. From there, the fit window $[t_{k_1},t_{k_2}]$ is extended adaptively, for as long as $|C_*^{ij}(t)|$ stays within a fraction $H$ of $|C_*^{ij}(t_{k_1})|\equiv C^{\rm ref}$, so that a single parameter $H>0$ (smaller $H$, tighter window) controls the fit for every coordinate pair $i,j$. 

We choose the optimal value of $H$ with a heuristic argument. For different sub-trajectories, and for a pair $i,j$, with the fitted parameters ($s_\ell,r_\ell,b_\ell^{(0)},\ldots,b_\ell^{(N_g)}$ for $\ell\le N_e$), we compute $D^{ij} = -\sum_\ell \dot B_\ell(t=0)$ and $\ddot C_*^{ij}(0^+) = \sum_\ell \ddot B_\ell(t=0)$. We then evaluate the traffic $\traffic$ for each sub-trajectory and check which $H$ yields the smallest variance. This value $H^\star$ is selected to extract the value $\traffic^\star=\traffic(H^\star)$ used to estimate $\sigma$, where $\traffic(H^\star)$ is the average of the traffic values of the sub-trajectories.

%Fig.~\ref{fig:C} illustrates three typical cases. For each, the left panel shows the full correlation function, and the right panel zooms in on its short-time behavior. Figs.~\ref{fig:C}(a),(b) show a case with $D^{ij}>0$ (illustrated by the dashed line, which is the linear part of the complete fit shown with the solid line) and a positive curvature $\ddot C_*(0^+)>0$, i.e.\ $\traffic<0$, compatible with equilibrium. Fig.~\ref{fig:C}(c),(d) still has $D^{ij}>0$ but a negative $\ddot C_*(0^+)<0$, i.e.\ $\traffic>0$, only possible out of equilibrium~\cite{diterlizzi2024vsrmodels}. In Fig.~\ref{fig:C}(e),(f) there is a case with $D^{ij}=0$, whose window ends on the rising side instead. All examples display points excluded from the fit at short times (blue diamonds) and at long times (green squares).

\begin{figure}[t]
\centering
\includegraphics[width=0.98\columnwidth]{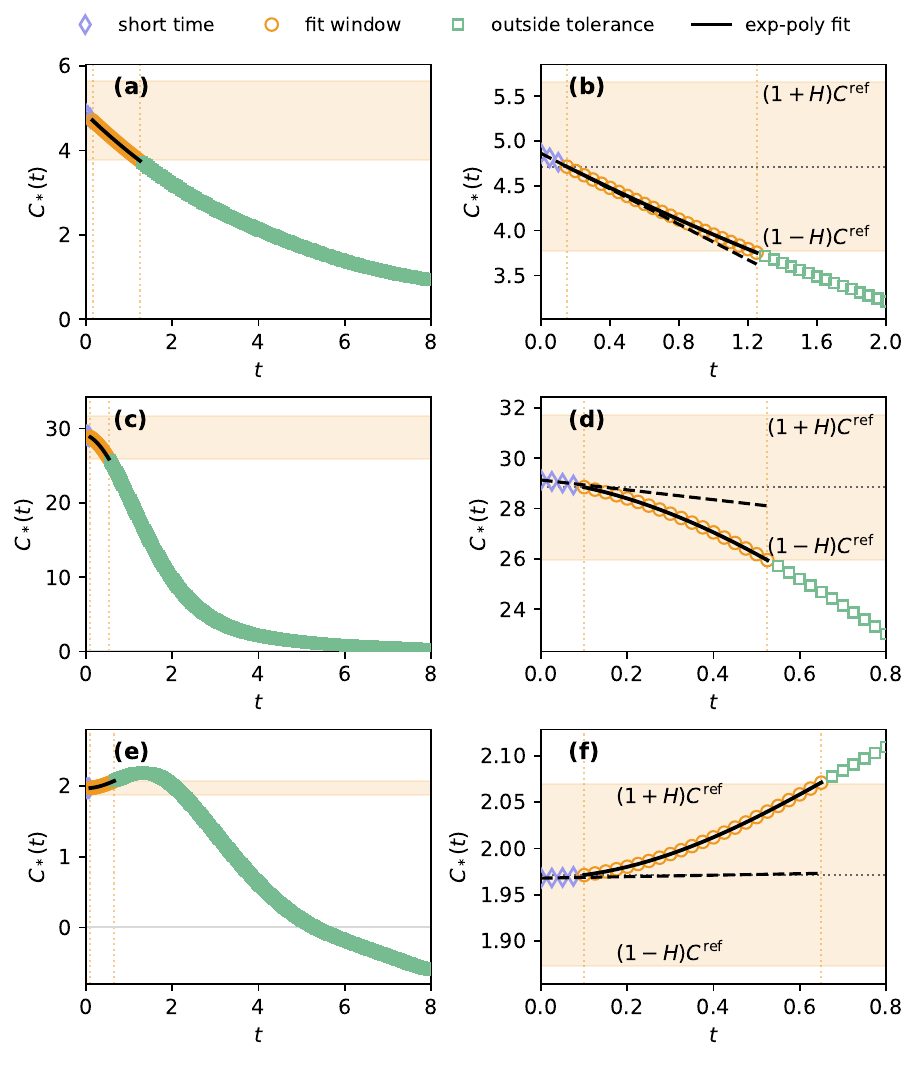}
\caption{Traffic-fit illustration for a generic element $C_*(t)\equiv C_*^{ij}(t)$: autocorrelations ($i=j$) in (a)-(d) and a cross-correlation ($i\neq j$) in (e)-(f). Symbols indicate each raw $C_*(t)$: excluded lags $k<k_1$ (diamonds), fit window $[t_{k_1},t_{k_2}]$ (circles), and later lags (squares). The reference $C^{\rm ref}\equiv C_*(t_{k_1})$ is compared to tolerance $H$; the shaded band $(1\pm H)C^{\rm ref}$ defines $t_{k_2}$. The black curve is the exp-poly fit~\eqref{exppoly}; its slope at $t=0$ (dashed) gives $\dot C_*(0^+)=-D^{ij}$. (a)-(b) equilibrium-like (left: full range; right: zoom on fit window), (c)-(d) nonequilibrium, (e)-(f) $D^{ij}=0$.}
\label{fig:C}
\end{figure}

Fig.~\ref{fig:C} illustrates three typical cases. For each, the left panel shows the full correlation function, and the right panel zooms in on its short-time behavior. Figs.~\ref{fig:C}(a),(b) show a case with $D^{ii}>0$ (illustrated by the dashed line, which is the linear part of the complete fit shown with the solid line) and a positive curvature $\ddot C_*^{ii}(0^+)>0$, i.e.\ $\traffic^i<0$, compatible with equilibrium. Fig.~\ref{fig:C}(c),(d) still has $D^{ii}>0$ but a negative $\ddot C_*^{ii}(0^+)<0$, i.e.\ $\traffic^i>0$, only possible out of equilibrium since $\sigma\ge4\traffic^i$ by Eq.~\eqref{lowerTS}. In Fig.~\ref{fig:C}(e),(f) there is a case with $D^{ij}=0$, whose window ends on the rising side instead. All examples display points excluded from the fit at short times (blue diamonds) and at long times (green squares).

%%%%%%%%%%%%%%%%%
%%%%%%%%%%%%%%%%%

\subsection{Computing the inflow rate}
\label{sec:practice-inflow}

Evaluating Eq.~\eqref{inflow} requires the steady-state score $\score(\xx)$, which we estimate on a spatial grid. We bin the trajectory's midpoints into cells with reference side $\Delta x$ along the first coordinate $x^1$ and proportional to $\Delta x$ along the other directions. With $\Gamma^i$ denoting side anisotropies ($\Gamma^1=1$), each cell volume is $\mathcal{V}=\Delta x^N\prod_i \Gamma^i$.
Then we estimate the local density $\rho$ from cell occupation counts (for cell $c$, with $n_c$ counts, $\rho_c=n_c/\mathcal{V}$), and obtain the score as the discrete gradient of $\ln\rho$ across neighboring cells.

The inflow rate $\inflow$ is then, from Eq.~\eqref{inflow}, the $\MD$-weighted, occupation-weighted covariance of this score field across all sufficiently occupied cells. Unlike the traffic term, this construction uses a spatial discretization, and its cost, both in memory and in the number of samples needed to populate every cell, grows exponentially with the dimension $N$.

Because a finite-sample, finite-resolution estimate of the score is intrinsically noisy, we additionally implement an optional Gaussian kernel smoothing of the binned fields before evaluating $\inflow$, see Ref.~\cite{cynar_github} for implementation details. We report both the raw and smoothed inflow-rate estimates for every example in Sec.~\ref{sec:examples}, since the choice of cell size $\Delta x$ affects them differently.

\subsection{Computing the local mean velocity}
\label{sec:practice-v2}

The $v^2$ method of Sec.~\ref{sec:v2method} is the other trajectory-based estimator of $\sigma$ considered in this work, and we use it as a baseline for the performance of the kinetic inference in the examples of Sec.~\ref{sec:examples}. To compare the two approaches on equal footing, we estimate $\velo(\xx)$ on the same spatial grid used for the inflow rate (Sec.~\ref{sec:practice-inflow}), so that $\sigma_{v^2}$ and $\inflow$ are computed on the same cells for a given cell size $\Delta x$. In a cell $c$ containing $n_c$ midpoints $(\xx_k+\xx_{k+1})/2$ of the trajectory, the local mean velocity is the sample average of the corresponding displacements per unit time~\cite{gonzalez2019experimental},
\begin{align}
\label{v-binned}
\velo(\xx_c) = \frac{1}{n_c}\sum_{k\in c}\frac{\xx_{k+1}-\xx_k}{dt} \, .
\end{align}
Binning at the midpoints corresponds to the Stratonovich convention, which is the one that yields the probability current. The estimate of $\sigma_{v^2}$ then follows by averaging $\velo\cdot\MD^{-1}\velo$ over the cells, weighted by their occupation, with the diffusion tensor obtained from Eq.~\eqref{Dhat}. As for the inflow rate, we also implement an optional Gaussian smoothing of the binned velocity field. Under partial observation, the same procedure applied to the recorded coordinates $\xx_\subs$, with $\MD_\subs$ obtained from their correlations through Eq.~\eqref{Dhat}, yields the marginal velocity $\tilde\velo$ and hence $\sigma_{v^2}^\subs$, which bounds $\sigma$ from below according to Eq.~\eqref{sigmav2S}.

The single-step displacements in Eq.~\eqref{v-binned} are dominated by the noise, whose contribution averages out only slowly with $n_c$. Each component of $\velo(\xx_c)$ thus carries a statistical error of order $\sqrt{2D/(n_c\,dt)}$, which is not small even for well-populated cells when $dt$ is small. Since Eq.~\eqref{sigmav2} is quadratic in $\velo$, these errors do not cancel but add up, producing a positive bias of order $N/(n_c\,dt)$ per cell. This bias is largest in poorly populated cells and grows as the cell size decreases, which explains the strong dependence of $\sigma_{v^2}$ on $\Delta x$ observed in Sec.~\ref{sec:examples}. The same bias affects $\sigma_{v^2}^\subs$ under partial observation, so that at small cell sizes the estimate can exceed the bound~\eqref{sigmav2S}, which holds for the exact marginal velocity.

\section{Examples}
\label{sec:examples}

We now benchmark CYNAR and the $v^2$ method on four systems chosen to probe different aspects of both estimators. For each, we compute the true $\sigma$ either from an analytical formula or from a long simulated trajectory using Eq.~\eqref{sigma-force-discr}. For the trajectory-based estimates, we typically analyze trajectories of $n=10^6$ points, split into five sub-trajectories to assess the variability of the estimates. Error bars denote the standard deviation of the estimates over the sub-trajectories, and the error on $4\traffic^\star+\inflow$ is obtained by combining those of $\traffic^\star$ and $\inflow$ in quadrature.

\subsection{Linear drift: the Brownian vortex}
\label{sec:linear}

%%%%%%%%%%%%%%%%%
\begin{figure}[t]
\centering
\includegraphics[width=\columnwidth]{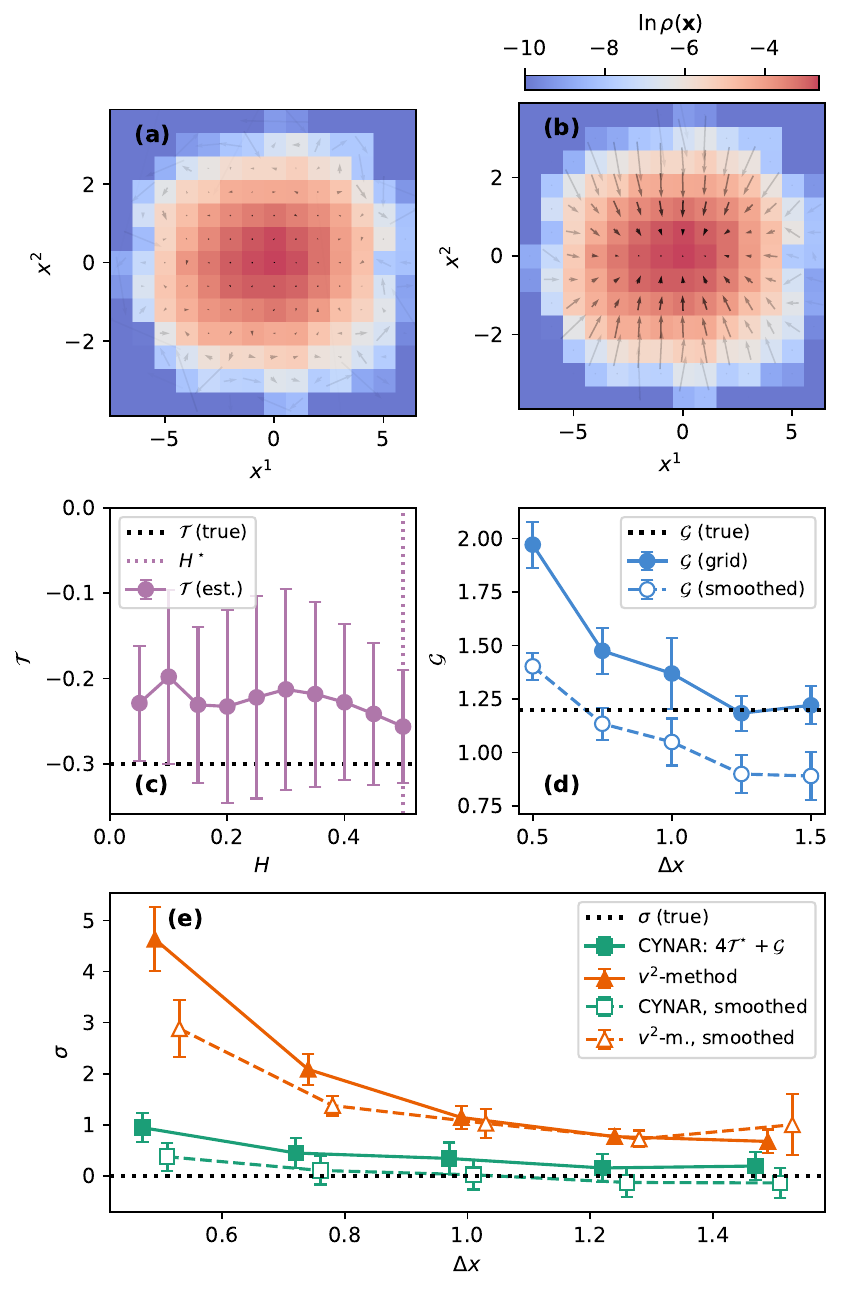}
\caption{Brownian vortex at the equilibrium point $\Phi=0$ (true $\sigma=0$). (a) Reconstructed local mean velocity $\velo(\bm x)$ and (b) score $\score(\bm x)$, overlaid as vector fields on the log-density $\ln\rho(\bm x)$, obtained with cell size $\Delta x=1$. (c) Traffic estimate versus the fit-window tolerance $H$ (Sec.~\ref{ssec:traffic}), used to select the most stable $\traffic^\star$. (d) Raw and smoothed inflow rate $\inflow(\Delta x)$ against its true value, versus the inflow-rate grid's cell size $\Delta x$. (e) CYNAR ($4\traffic^\star+\inflow$) versus the $v^2$ method, both raw and smoothed, against the true $\sigma$, versus $\Delta x$.}
\label{fig:linear-eq}
\end{figure}
%%%%%%%%%%%%%%%%%

%%%%%%%%%%%%%%%%%
\begin{figure}[t]
\centering
\includegraphics[width=\columnwidth]{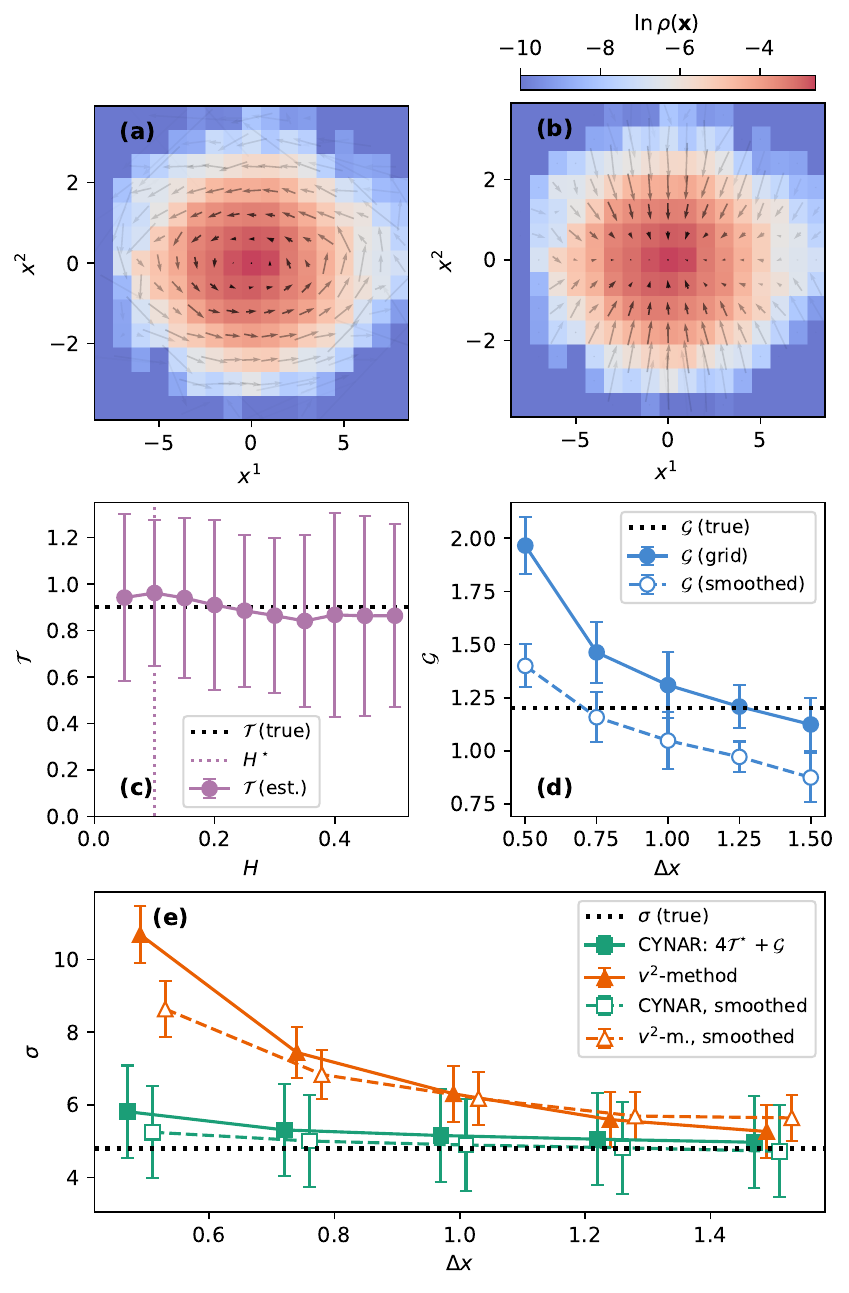}
\caption{As in Fig.~\ref{fig:linear-eq} but for the Brownian vortex at a nonequilibrium point $\Phi=2$ (true $\sigma=4.8$).}
\label{fig:linear-neq}
\end{figure}
%%%%%%%%%%%%%%%%%

We first consider the Brownian vortex model studied in Ref.~\cite{diterlizzi2024vsrmodels}. It is a linear stochastic model describing a particle in an anisotropic harmonic trap $U(x,y)=\kappa(\alpha x^2+y^2)/2$, with $0<\alpha<1$ setting the anisotropy between the two coordinates $x\equiv x^1$ and $y\equiv x^2$, and driven by a nonconservative force $(-y \Phi,\alpha x \Phi)$ perpendicular to $-\nabla U$. This represents a vortex circulating the equipotential ellipses and gives a drift
\begin{equation}
\drift(\xx)=\MA\xx\qquad\textrm{with}\qquad \MA=\begin{pmatrix}-\alpha\kappa & -\Phi\\ \alpha\Phi & -\kappa\end{pmatrix}.
\end{equation}
Here, for simplicity, we choose to have isotropic diffusion $\MD=T\,{\mathbb I}$ and set $\kappa=T=1$ and $\alpha=1/5$, leaving the nonequilibrium drive $\Phi$ as the only free parameter. A peculiarity of this Brownian vortex is that, even out of equilibrium, the stationary density keeps the same Gaussian shape it would have at $\Phi=0$, $\rho(x,y)\propto\exp(-\kappa(\alpha x^2+y^2)/2T)$. The closed forms of $\sigma$, $\traffic$, and $\inflow$ for this linear model with $\kappa=T=1$ are~\cite{diterlizzi2024vsrmodels}
\begin{equation}
  \sigma=(1+\alpha)\Phi^2, \quad \traffic=\frac{(1+\alpha)(\Phi^2-1)}{4}, \quad \inflow=1+\alpha.
\end{equation}

We illustrate this on two representative points, both at $\alpha=0.2$: the equilibrium point $\Phi=0$ (Fig.~\ref{fig:linear-eq}) and a nonequilibrium point $\Phi=2$ (Fig.~\ref{fig:linear-neq}), each summarized in the same five-panel layout. The anisotropy due to $\alpha=0.2$ suggests using rectangular cells with sides $\Delta x$ and $0.6\,\Delta x$ for binning.

Fig.~\ref{fig:linear-eq}(a) shows the vector field of the reconstructed mean velocity $\velo(\bm x)$ on top of the log-density $\ln\rho(\bm x)$, at the cell size $\Delta x=1$. At this equilibrium point $\Phi=0$, the reconstructed $\velo$ is essentially null as expected, but small statistical fluctuations will bias the $v^2$ method. Fig.~\ref{fig:linear-eq}(b) shows the score $\score(\bm x)$, which is, of course, nonzero also in equilibrium, where it is proportional to the conservative force, $\score=-\nabla U/T$.

Fig.~\ref{fig:linear-eq}(c) plots the traffic estimate as a function of the fit-window tolerance $H$, highlighting $H^\star$ where the estimate is most stable over independent trajectories and where we get $\traffic^\star=-0.26\pm0.07$, compatible with the true $\traffic=-0.3$. In Fig.~\ref{fig:linear-eq}(d), we show the reconstruction of raw and smoothed inflow rate $\inflow(\Delta x)$ for cell size $\Delta x \in [0.5,1.5]$. Both estimates are close to the true value $\inflow=1.2$ and decrease as $\Delta x$ grows, which is expected for an information-theoretic quantity estimated in coarser partitions. For a similar reason, information loss from smoothing also yields smaller values at the same $\Delta x$.
Fig.~\ref{fig:linear-eq}(e) combines these into $\sigma=4\traffic^\star+\inflow(\Delta x)$ versus $\sigma_{v^2}(\Delta x)$, raw and smoothed, and also shows the true $\sigma=0$. The CYNAR estimate relaxes to zero from a much smaller starting value and faster than the $v^2$ method. This is consistent with CYNAR being both stable and unbiased (it can fluctuate on both sides of zero, unlike the non-negative estimate of the $v^2$ method).

The same five panels at the nonequilibrium point $\Phi=2$ (Fig.~\ref{fig:linear-neq}) lead to similar conclusions. Fig.~\ref{fig:linear-neq}(a) now shows a clear circulating current connected to broken detailed balance, while the score in Fig.~\ref{fig:linear-neq}(b) remains the same as in Fig.~\ref{fig:linear-eq}(b), as expected for the Brownian vortex. We continue using the cell size $\Delta x=1$ (since the stationary density is $\Phi$-independent in the Brownian vortex). Fig.~\ref{fig:linear-neq}(c) gives $\traffic^\star=0.96\pm0.32$ at $H^\star=0.10$, against the true $\traffic=0.9$, while Fig.~\ref{fig:linear-neq}(d)-(e) show the inflow-rate and CYNAR estimates staying close to the true $\inflow$ and $\sigma$ throughout the $\Delta x$ range. However, Fig.~\ref{fig:linear-neq}(e) shows that the $v^2$ method drifts substantially, approaching the correct value only at the largest cell sizes tested.
%%%%%%%%%%%%%%%%%
\begin{figure}[t]
\centering
\includegraphics[width=0.98\columnwidth]{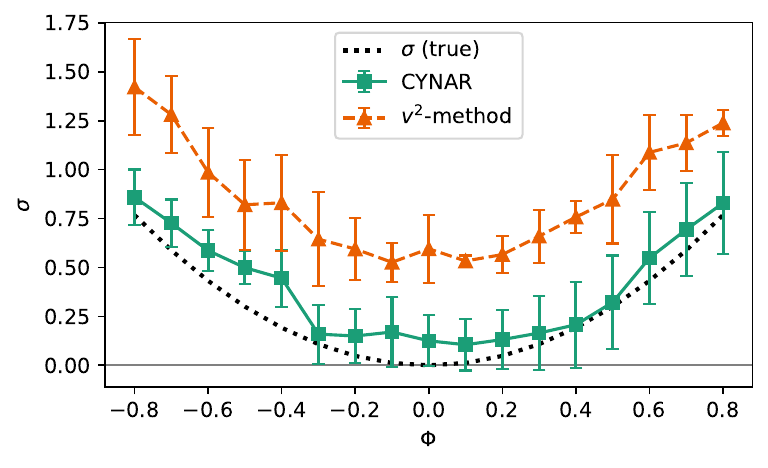}
\caption{For the Brownian vortex, CYNAR and $v^2$ method estimates of $\sigma$ (both using the smoothed inflow rate/local mean velocity) vs.~the nonequilibrium strength $\Phi$, compared to the exact $\sigma(\Phi)$ (dotted). We use $\Delta x = 1$ and the same parameters as in previous figures.}
\label{fig:linear-Phi}
\end{figure}
%%%%%%%%%%%%%%%%%

To further check CYNAR's unbiasedness, we vary $\Phi$ around the equilibrium point, as shown in Fig.~\ref{fig:linear-Phi}. CYNAR's estimates are close to the true $\sigma(\Phi)$, while the $v^2$ method shows positive bias.

%%%%%%%%%%%%%%%%%
\begin{figure}[t]
\centering
\includegraphics[width=0.98\columnwidth]{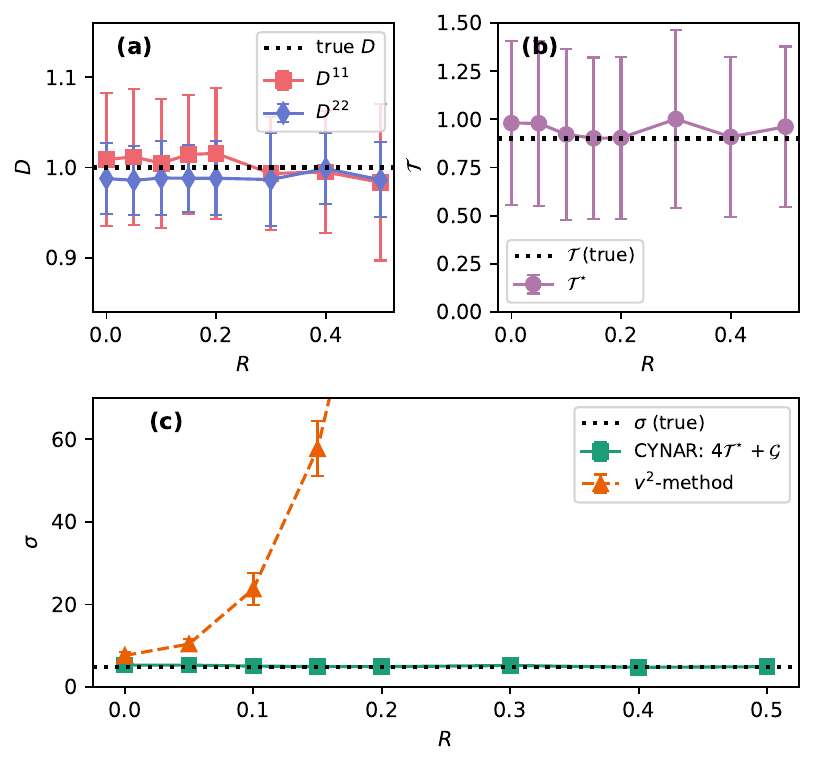}
\caption{For the Brownian vortex, robustness to measurement noise as a function of the relative noise strength $R$: (a) reconstructed diffusion coefficients $D^{11}, D^{22}$ stay close to the true value (not shown, $D^{12}\simeq 0$), (b) correspondingly, the traffic estimate $\traffic^\star$ stays close to $\traffic_{\rm true}=0.9$, and (c) $\sigma_{\rm CYNAR}=4\traffic^\star+\inflow$ stays close to $\sigma_{\rm true}=4.8$, while the $v^2$ method grows quickly with $R$.}
\label{fig:linear-noise}
\end{figure}
%%%%%%%%%%%%%%%%%

To probe the robustness to measurement noise, we use the reference point $\alpha=0.2$, $\Phi=2$ and analyze the corrupted time series $q^i=x^i+X^i\,R\,\psi^i$. Here, $X^i=\sqrt{\mathrm{Var}(x^i)}$ is the typical fluctuation of $x^i$, and $\psi^i$ is a standard Gaussian white noise, independent for each component and time step. We vary the relative noise strength $R$ from $0$ to $0.5$, where the added noise reaches half of the typical fluctuation of each component. Since this noise is uncorrelated in time, it only affects the correlation functions at $t=0$, which the traffic fit excludes by using lags $t\ge t_{k_1}>0$. Hence, both the reconstructed diffusion coefficients (Fig.~\ref{fig:linear-noise}(a)) and the traffic estimate (Fig.~\ref{fig:linear-noise}(b)) stay close to their true values over the tested range of $R$, and $\sigma=4\traffic^\star+\inflow$ stays close to $\sigma_{\rm true}$ even for $R=0.5$ (Fig.~\ref{fig:linear-noise}(c)). The estimates based on the local mean velocity are instead strongly affected by the added noise and quickly increase with $R$, reaching values orders of magnitude above the true one, as shown in Fig.~\ref{fig:linear-noise}(c). This robustness to measurement noise confirms the findings of Ref.~\cite{diterlizzi2025force}.

\subsection{Hair-bundle model}
\label{sec:hair}

\begin{figure}[t]
\centering
\includegraphics[width=\columnwidth]{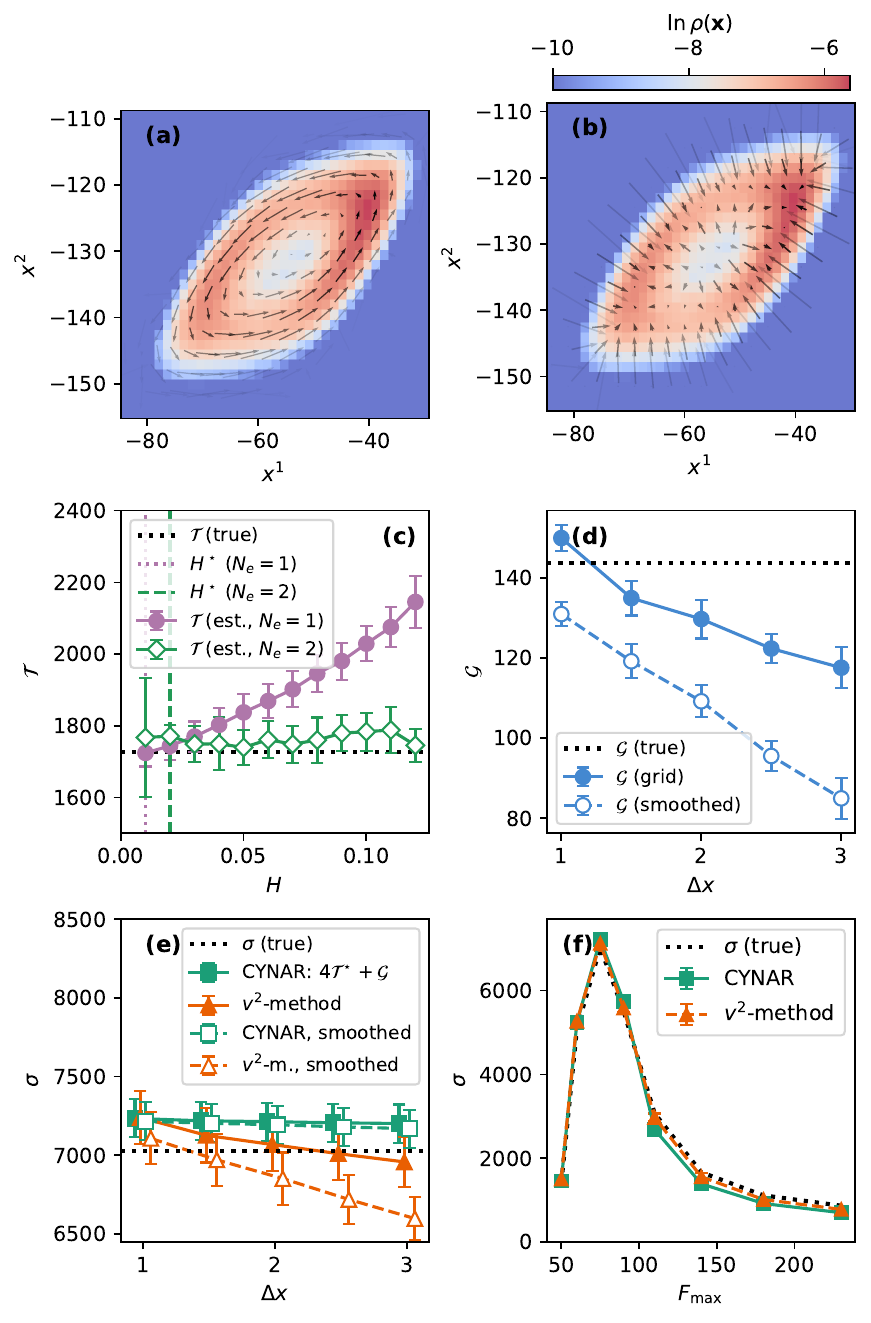}
\caption{Hair-bundle model, with the parameter set given in Sec.~\ref{sec:hair}. (a)-(e) As in Fig.~\ref{fig:linear-eq}. In (c), the traffic estimate is shown both for the standard choice of $N_e=1$ (filled circles) and $N_e=2$ (open diamonds), each with its own $H^\star$; $N_e=1$ never plateaus over the tested range, while $N_e=2$ converges close to the true value and is used for the following panels.
(f) CYNAR and $v^2$ method estimates of $\sigma$, obtained with cell size $\Delta x=1.5$, vs the motor-force amplitude $F_{\rm max}$. Both follow the true $\sigma(F_{\rm max})$ (dotted).}
\label{fig:hair-summary}
\end{figure}

As a first nonlinear test case, we use the two-state gating-spring model of Nadrowski et al.~for the bullfrog's saccular hair bundle \cite{hair0,hair4}, an actively driven mechanosensory system known to exhibit spontaneous oscillations. Its two degrees of freedom are the stereociliary bundle position $x$ and the position $y$ of the adaptation motors (i.e.~again $x^1\equiv x$, $x^2\equiv y$), coupled through an $N_{\rm ch}$-element two-state transduction channel whose open probability $P_0(z)$, a function of the gate extension $z=x-y$, sets the forces
\begin{align}
P_0(z) &= \left(1+\exp\!\left(-\frac{k_{gs}d\,z}{N_{\rm ch}T}+\frac{\Delta G+k_{gs}d^2/2N_{\rm ch}}{T}\right)\right)^{-1}, \\
F^x &= -k_{gs}z-k_{sp}x+k_{gs}d\,P_0(z), \\
F^y &= k_{gs}z-k_{gs}d\,P_0(z)-F_{\rm max}\big(1-S\,P_0(z)\big),
\end{align}
with overdamped dynamics $\dot x=\mu_1F^x+\sqrt{2\mu_1T}\,\xi^1$, $\dot y=\mu_2F^y+\sqrt{2\mu_2T_{\rm eff}}\,\xi^2$ (i.e., in our notation, $\drift(\xx)=(\mu_1F^x,\mu_2F^y)$ and $\MD$ with elements $D^{11}=\mu_1T$, $D^{22}=\mu_2T_{\rm eff}$, $D^{12}=D^{21}=0$), where the motor coordinate $y$ is driven at an effective temperature $T_{\rm eff}=1.5\,T$, modeling the extra fluctuations injected by myosin-motor activity. This, together with the active-force term $F_{\rm max}(1-SP_0(z))$ in $F^y$, is what drives the system out of equilibrium. We start by using the parameter values of Ref.~\cite{hair4}: $S=1$, $F_{\rm max}=75$, $k_{gs}=0.75$, $k_{sp}=0.60$, $d=61$, $\Delta G=10\, T$, $N_{\rm ch}=50$, $\mu_1=357.1$, $\mu_2=100.0$, $T=4.11$, in units of nm, pN, and s (so e.g.\ $\mu_1$ is in nm/(pN$\cdot$s)). Since no closed form is available, we obtain the reference values of $\sigma$, $\traffic$, and $\inflow$ as described in Sec.~\ref{sec:drift}, from a long ($10^6$-step, $dt=10^{-4}$) trajectory. At the reference point $F_{\rm max}=75$, $S=1$, this gives $\sigma\simeq6993\,k_{\rm B}/\mathrm{s}$ and $\traffic\simeq1718\,k_{\rm B}/\mathrm{s}$.  Fig.~\ref{fig:hair-summary} summarizes our findings.

%%%%%%%%%%%%%%%%%
%%%%%%%%%%%%%%%%%

Fig.~\ref{fig:hair-summary}(a) shows the reconstructed local mean velocity $\velo(\bm x)$ on top of the log-density $\ln\rho(\bm x)$, estimated with square cells of size $\Delta x =1.5$. The oscillatory limit cycle of the hair bundle appears as a closed circulation, similar to the rotation in the Brownian vortex. Fig.~\ref{fig:hair-summary}(b) instead shows the score $\score(\bm x)$, which, as usual, points toward the maxima of $\rho(\xx)$.

Here we also fit the traffic with a richer, two-term exp-poly model ($N_e=2$). Fig.~\ref{fig:hair-summary}(c) shows that with the single-component fit ($N_e=1$, filled circles), $\traffic(H)$ never plateaus over the tested range of $H$. This behavior signals that this fit might not consistently report the system's short-time correlation shape. The fit with two components ($N_e=2$) instead gives a flat $\traffic(H)$ and is used for all other panels in this figure. We describe this step in detail because it illustrates how the fit is chosen without knowing the true $\traffic$. Interestingly, the minimum-variance estimate for $N_e=1$ is close to that for $N_e=2$ and even slightly closer to the true value, which, however, one could not know without a reference.

Fig.~\ref{fig:hair-summary}(d) shows the raw and smoothed reconstructed inflow rate $\inflow(\Delta x)$ for several cell sizes, together with the true value $\inflow=142\,k_{\rm B}/\mathrm{s}$. Here, smaller cell sizes yield a more precise estimate of $\inflow$, while larger ones underestimate it. In contrast, the $v^2$ estimate of $\sigma$ (Fig.~\ref{fig:hair-summary}(e), both raw and smoothed versions) approaches the true value for larger $\Delta x$. Increasing $\Delta x$ further would drive the estimate to zero, so we still have no blind way to choose the best $\Delta x$. The estimates of $\sigma$ with CYNAR in Fig.~\ref{fig:hair-summary}(e) are stable and of comparable precision to those of the $v^2$ method. This low-dimensional, far-from-equilibrium system can thus be equally resolved with both methods.
The estimates obtained by varying $F_{\rm max}$ to sample the spontaneously oscillating regime ($F_{\rm max}\gtrsim 50$, see Fig.~\ref{fig:hair-summary}(f)) confirm that both methods predict $\sigma$ with good precision.  

The case in which only the bundle position $x$ is observed, as in experiments, was analyzed in Ref.~\cite{diterlizzi2025force}. There, the traffic of $x$ recovered roughly half of $\sigma$ in the most active regimes, a dissipation about three orders of magnitude larger than the estimates based on time irreversibility~\cite{roldan2021quantifying,ghosal2022inferring}, whereas the $v^2$ method gives zero when a single coordinate is observed.

\subsection{Stochastic Lorenz model}
\label{sec:lorenz}

In this section, we test both the lower bounds in Sec.~\ref{sec:bound} and show that CYNAR may provide valuable information even when observing a single coordinate.
To this end, we choose to analyze a model with $N=3$ degrees of freedom that sits between the two-dimensional examples above and the high-dimensional Lorenz-96 model of Sec.~\ref{sec:lorenz96}. 
Specifically, we consider the classic Lorenz system~\cite{lorenz1963deterministic}, $\dot x=s(y-x)$, $\dot y = rx-y-zx$, $\dot z = xy-bz$, with additive noise. Its drift has constant divergence, $\nabla\cdot\drift=-(s+1+b)$, so that Eq.~\eqref{inflow-force} gives the true inflow rate $\inflow_{\rm true}=s+1+b$.

\begin{figure}[t]
\centering
\includegraphics[width=0.98\columnwidth]{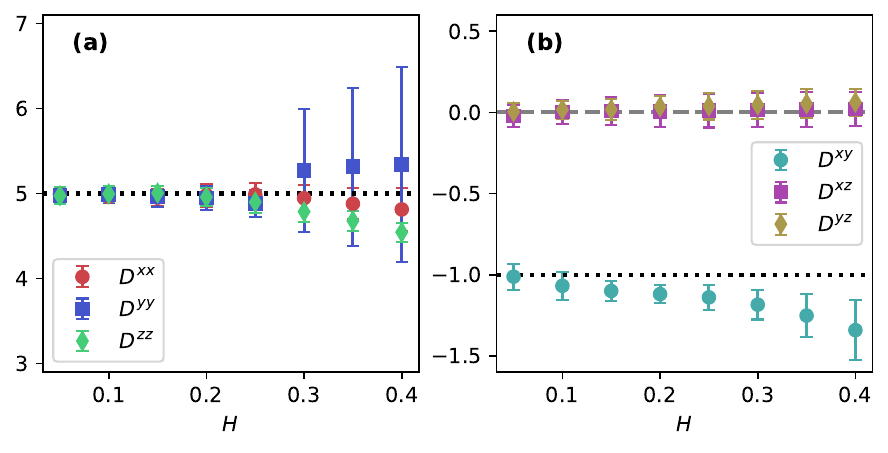}
\caption{Fitted elements of the Lorenz model's diffusion tensor $\MD$ versus the fit-window tolerance $H$: (a) Diagonal components $D^{xx}$, $D^{yy}$, $D^{zz}$, and their true value $5$ and (b) off-diagonal components $D^{xy}$, $D^{xz}$, $D^{yz}$, with, respectively, true values $-1$, $0$, and $0$ (horizontal lines).}
\label{fig:LorD}
\end{figure}

\begin{figure}[tb]
\centering
\includegraphics[width=0.98\columnwidth]{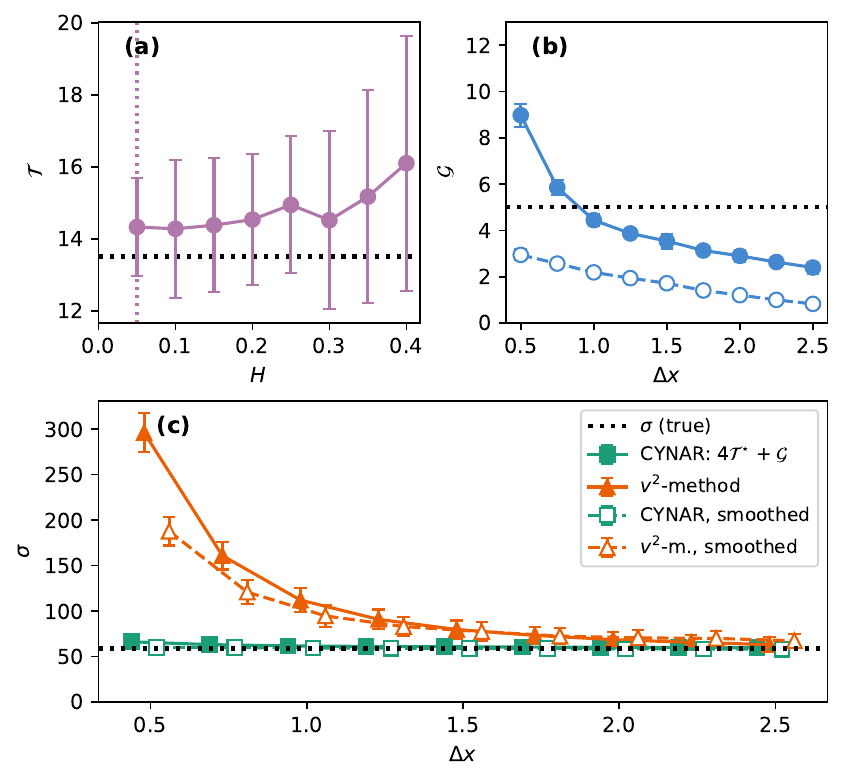}
\caption{Lorenz model ($N=3$): same as in Fig.~\ref{fig:linear-neq}(c)-(e).}
\label{fig:Lor}
\end{figure}

Throughout this subsection we use $r=10$, $s=3$, $b=1$ and a non-diagonal diffusion tensor,
\begin{align}
  \MD=\left(\begin{matrix}5&-1&0\\-1&5&0\\0&0&5\end{matrix}\right)
\end{align}
to check whether the fit of the correlation functions recovers all of its elements. Fig.~\ref{fig:LorD}(a) shows the three diagonal components $D^{xx},D^{yy},D^{zz}$ and (b) the three off-diagonal ones $D^{xy},D^{xz},D^{yz}$, both as a function of the fit-window tolerance $H$. All six track their true values reasonably closely, and the traffic-vs-$H$ sweep in Fig.~\ref{fig:Lor}(a) also shows a precise prediction.

%%%%%%%%%%%%%%%%%
%%%%%%%%%%%%%%%%%

At small cell sizes, the inflow rate is estimated fairly well (Fig.~\ref{fig:Lor}(b)). However, these cell sizes are too small for the $v^2$ method, which strongly overestimates $\sigma$ there (Fig.~\ref{fig:Lor}(c)). This $\sigma$ estimate, both raw and smoothed, starts nearly an order of magnitude too high at the smallest cell size and only slowly relaxes toward a fairly reasonable $\sigma$ estimate as $\Delta x$ grows, yet remains too high. Fig.~\ref{fig:Lor}(c) also shows that, again, the CYNAR estimate barely changes with the cell size $\Delta x$ and stays close to the true $\sigma$ in the tested range. This suggests that CYNAR performance does not degrade with system dimensionality as quickly as the $v^2$ method, as we confirm in higher dimensions in Sec.~\ref{sec:lorenz96}. As anticipated in Sec.~\ref{sec:bound}, the large positive traffic dominates far from equilibrium, while the smaller inflow rate, whose histogram-based evaluation degrades with $N$, is marginal.

%%%%%%%%%%%%%%%%%
%%%%%%%%%%%%%%%%%

To test the inference scheme under partial observation, we apply both estimators to a subset $\subs$ of the coordinates, using only the trajectory of $\xx_\subs$ and the block $\MD_\subs$ of the diffusion tensor estimated from its correlations. Since the Lorenz drift couples all three coordinates nonlinearly, observing any proper subset of $(x,y,z)$ leaves part of the dynamics hidden. By Eq.~\eqref{lowerTS}, the partial CYNAR estimate $\max(4\traffic^\subs+\inflow^\subs,0)$ bounds from below $\sigma^\subs$, Eq.~\eqref{sigmaS}, and hence $\sigma$, for any $\MD$. Since the inflow rate is much lower than $\sigma$ in this example, the partial CYNAR estimate is expected to be close to $\sigma^\subs$ (see Sec.~\ref{sec:v2method}).

%%%%%%%%%%%%%%%%%
\begin{figure}[t]
\centering
\includegraphics[width=0.98\columnwidth]{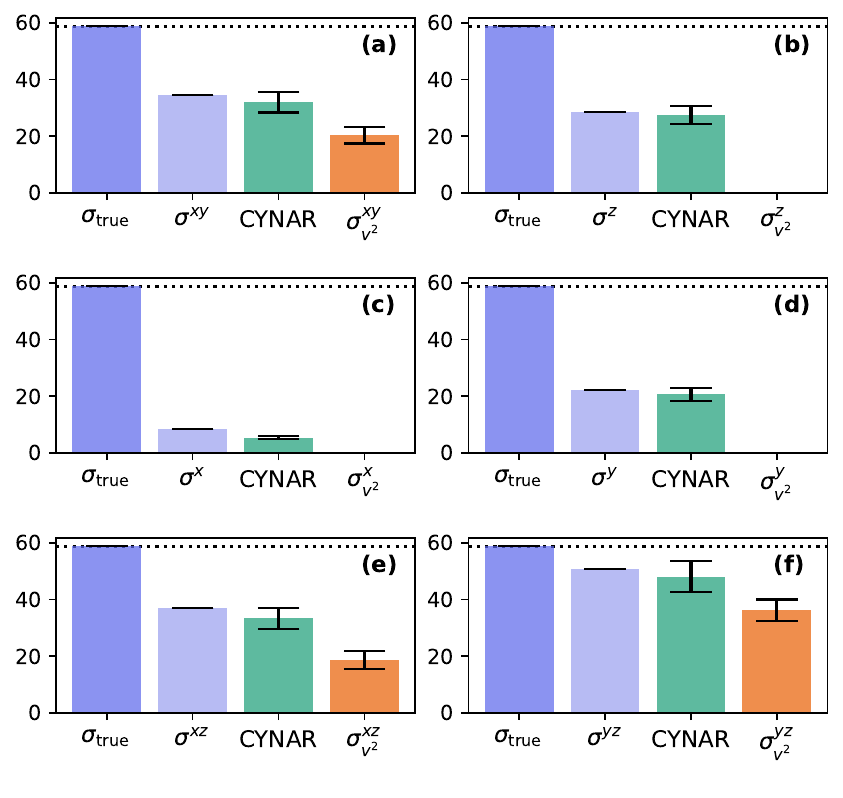}
\caption{Partial observation of the Lorenz model for six observed subsets $\subs$. Each panel shows $\sigma_{\rm true}$ (also as a dotted line), the dissipation $\sigma^\subs$ associated with the observed subset, Eq.~\eqref{sigmaS} (evaluated with the $v^2$ method on the full trajectory with the known $\MD$, hence slightly biased upward), and the estimates obtained from the observed coordinates with CYNAR, $4\traffic^{\subs\star}+\inflow^\subs$, and with the $v^2$ method. (a) $\{x,y\}$ and (b) $\{z\}$, which do not share noise with the hidden coordinates, so that $\sigma_{\rm true}=\sigma^{xy}+\sigma^z$. (c) $\{x\}$ and (d) $\{y\}$, single coordinates coupled to a hidden one through the noise ($D^{xy}\neq0$). (e) $\{x,z\}$ and (f) $\{y,z\}$, pairs with diagonal $\MD_\subs$ that share noise with a hidden coordinate. The CYNAR estimate, a lower bound on $\sigma^\subs$ according to Eq.~\eqref{lowerTS}, stays close to it in all cases, whereas the $v^2$ estimate relies on the marginal velocity and vanishes when a single coordinate is observed.}
\label{fig:Lor-partial}
\end{figure}
%%%%%%%%%%%%%%%%%

To probe how the noise structure affects partial observation, we consider the six observed subsets shown in Fig.~\ref{fig:Lor-partial}. For $\subs=\{x,y\}$ and $\subs=\{z\}$, panels (a) and (b), the observed coordinates do not share noise correlations with the hidden ones, since $D^{xz}=D^{yz}=0$. The entropy production rate then splits into the contributions of the two complementary subsets, $\sigma_{\rm true}=\sigma^{xy}+\sigma^z$, where $\sigma^{xy}$ and $\sigma^z$ denote $\sigma^\subs$ for $\subs=\{x,y\}$ and $\subs=\{z\}$, see Eq.~\eqref{sigmaS}. For the other subsets, $\subs=\{x\}$, $\{y\}$, $\{x,z\}$, and $\{y,z\}$ in panels (c)-(f), an observed coordinate is coupled to a hidden one through the noise, $D^{xy}=-1\neq0$. In this case, $\sigma$ does not split into contributions of observed and hidden coordinates, but $\sigma^\subs$ remains well defined for each subset and Eq.~\eqref{lowerTS} still applies, since it holds for any positive-definite $\MD$.

In all six cases, the CYNAR estimate $4\traffic^{\subs\star}+\inflow^\subs$ stays close to $\sigma^\subs$, while the $v^2$ method lies below it, as expected from Eq.~\eqref{sigmav2S}, since it only accesses the marginal velocity of the observed coordinates (see Sec.~\ref{sec:v2method}). When a single coordinate is observed, as for $\{x\}$, $\{y\}$, and $\{z\}$, the $v^2$ method gives zero regardless of the true dissipation, whereas the CYNAR estimate obtained from the correlations and the histogram of the observed coordinate remains positive and provides a nontrivial lower bound on $\sigma^\subs$. 
Finally, for $\{x,z\}$ and $\{y,z\}$ the block $\MD_\subs$ is diagonal, so that $\sigma^\subs$ is the sum of the single-coordinate contributions, e.g., $\sigma^{xz}=\sigma^x+\sigma^z$, as seen by comparing panels (b), (c), and (e), in agreement with the partition argument behind Eq.~\eqref{lowerT}.

\subsection{Stochastic Lorenz-96 model}
\label{sec:lorenz96}

Finally, we use the Lorenz-96 model \cite{lorenz1996predictability}, $\dot x^i = (x^{i+1}-x^{i-2})x^{i-1}-x^i+F$ (indices mod $N$) with additive noise. Its $N$ degrees of freedom allow us to probe the high-dimensional case. As for the Lorenz model, the drift's divergence is constant, $\nabla\cdot\drift=-N$, since $\partial_i a^i=-1$ for every $i$, so Eq.~\eqref{inflow-force} again gives the true inflow rate $\inflow_{\rm true}=N$, which grows only linearly with $N$.

\begin{figure}[t]
\centering
\includegraphics[width=\columnwidth]{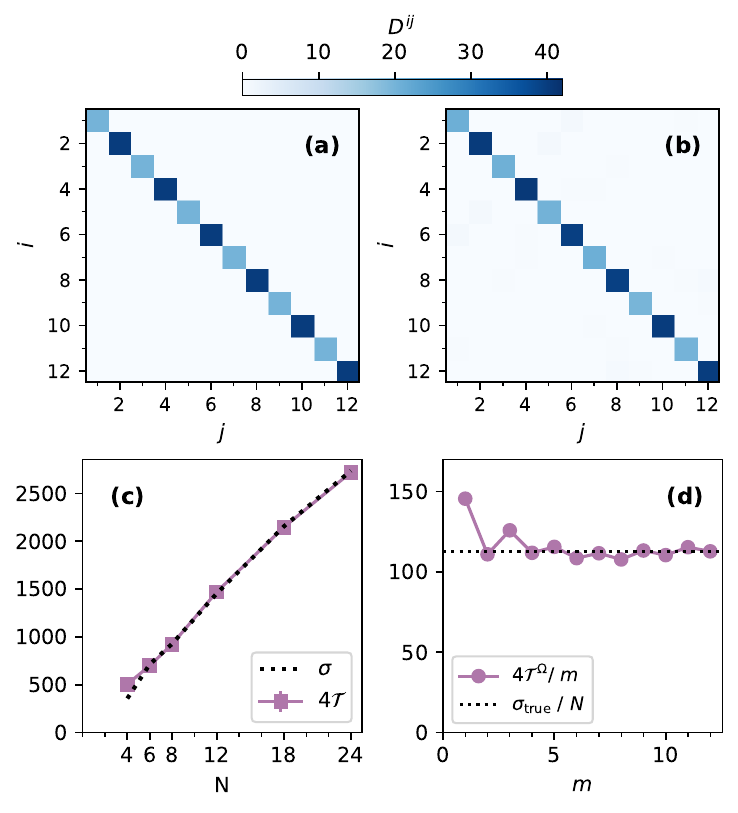}
\caption{For the Lorenz-96 model with $N=12$, with a diagonal $\MD$ of two alternating values, (a) true $\MD$ and (b) $\MD$ recovered from the correlation-curvature fit used for $\traffic$. (c) CYNAR's traffic-only estimate $4\traffic$ and true $\sigma$ as a function of $N$. (d) Per-site running average $4\traffic^\subs/m$ versus the number $m$ of observed contiguous sites and global per-site true $\sigma_{\rm true}/N$ (dotted line).}
\label{fig:L-96}
\end{figure}

With, say, 20 bins per axis, a spatial grid for $\inflow$ or $\sigma_{v^2}$ needs a number of cells growing as $20^N$, which quickly becomes infeasible. We therefore rely here on the traffic-only bound $\max(4\traffic,0)$, i.e., Eq.~\eqref{lowerTS} with $\subs$ including all degrees of freedom, which only requires pairwise correlation fits with a numerical cost $\sim N^2$. We use a diagonal $\MD$ with two alternating values, $D^{ii}=D_0$ for odd $i$ and $2D_0$ for even $i$, with $D_0=20$, so that the noise level is well defined at every $N$ and the recovered $\MD$ is easy to inspect visually. At a representative $N=12$, Fig.~\ref{fig:L-96}(a)-(b) shows that the true $\MD$ (panel (a)) is recovered from the correlation fit used for $\traffic$ (panel (b)). Fig.~\ref{fig:L-96}(c) then shows, for $N=4$ to $N=24$ with $F=8$, that $4\traffic$ closely tracks $\sigma_{\rm true}$ because the system is far from equilibrium and the true inflow rate $\lesssim\sigma/100$ at every tested $N$.

%%%%%%%%%%%%%%%%%
%%%%%%%%%%%%%%%%%

In Fig.~\ref{fig:L-96}(d), we show results on partial observation of degrees of freedom. Beyond the lower bound $\sigma\ge\max(4\traffic,0)$ based on full traffic, we test the hypothesis that observing even a handful of coordinates in a high-dimensional system may provide a rescaled estimate of $\sigma$ when the system is sufficiently homogeneous. Indeed, when we observe a growing contiguous block of $m$ sites ($m\le N=12$) and track the per-site average $4\traffic^\subs/m$, we find that it is reasonably close to the true per-site dissipation rate $\sigma_{\rm true}/N$, even when a single site is observed ($m=1$).

\section{Discussion}
\label{sec:discussion}

Like any inference scheme, CYNAR relies on a few assumptions that are worth keeping in mind when analyzing experimental data. The dynamics is assumed to be overdamped, with a constant diffusion tensor and a well-defined steady state, and the sampling time $dt$ should be short enough to resolve the short-time curvature of the correlation functions. Inertial effects or temporally correlated noise may affect this short-time behavior and, with it, the estimates of $\MD$ and $\traffic$. Under partial observation, the bound~\eqref{lowerTS} cannot exceed the dissipation $\sigma^\subs$ associated with the observed coordinates, Eq.~\eqref{sigmaS}, which may be considerably smaller than $\sigma$ when much of the dissipation involves hidden degrees of freedom. Even in this case, a positive estimate provides evidence that the system is out of equilibrium.

Turning to the practical aspects that emerged from our examples, we defined reasonably robust and reliable criteria for automatically choosing the traffic fitting window, although this choice remains system-specific. The exp-poly model in Sec.~\ref{ssec:traffic} should be fit over a time interval short enough to capture the short-time curvature of the correlation function, yet long enough to remain numerically stable, and we did not find a universally valid selection. 
In practice, we determine the window empirically, computing the traffic from independent sub-trajectories and picking the fitting interval that yields the smallest fluctuations, which defines $\traffic^\star$. This procedure is sensible but not necessarily optimal, and additional tests may suggest better approaches.

Our examples also show that selecting the optimal cell size and smoothing amount is nontrivial. When computing the score $\score(\xx)$ or the local velocity $\velo(\xx)$, grid-based estimators require care because the cell size determines the results, and no criterion has emerged as reliable for determining the best $\Delta x$. In particular, neither the inflow rate nor $\sigma_{v^2}$ settles on a plateau as $\Delta x$ varies. At small $\Delta x$, a cell adjacent to a poorly visited neighbor can receive a spurious, large score value from the finite-difference log-density gradient, a discretization artifact whose magnitude shrinks with cell size. Similarly, averaging steps with midpoint in a given cell yields an estimate of $\velo$ that becomes statistically more accurate with growing $\Delta x$, but risks averaging over too wide regions. In the limit of large $\Delta x$, the discretization bias of a coarser histogram then leads to estimates below the true values. The CYNAR estimate of $\sigma$ is less affected by these issues, since the traffic does not depend on $\Delta x$ and the inflow rate is typically low compared with $4\traffic$ far from equilibrium.

Smoothing $\rho(\xx)$ and/or the vector fields (with bandwidth tied to the cell size; see the CYNAR repository~\cite{cynar_github} for implementation details) mitigates the problem in the small $\Delta x$ regime. This may, however, be detrimental for a thin or curved state-space support, where averaging in unvisited neighbors dilutes the smoothed estimate. In some examples the additional averaging was excessive, removing fine details of $\rho(\xx)$ and producing an overly low estimate of $\inflow$. A posteriori, we conclude that smoothing $\rho,\velo,$ and $\score$ was not essential in our examples and that no clear improvement in the inflow-rate estimate supported the effort to find an optimal smoothing protocol. It remains to be studied whether smoothing can compensate for poor data acquisition, with short time series that do not allow histograms to be filled satisfactorily.

\section{Conclusions}
\label{sec:conclusions}

Our tests of the kinetic inference of entropy production introduced in Ref.~\cite{diterlizzi2025force}, here implemented in the CYNAR pipeline, reveal several appealing features for estimating $\sigma$ from trajectories without force measurements. Compared with the $v^2$ method, CYNAR is often more accurate, more stable under changes in spatial discretization, robust to measurement noise (as also tested in Ref.~\cite{diterlizzi2025force}), and more informative under partial observation. In addition, we have shown that the same quantities, evaluated on any subset of observed coordinates, bound $\sigma$ for any constant, positive-definite diffusion tensor and without knowledge of the hidden degrees of freedom. The traffic already provides such a bound, which remains available in high-dimensional systems where the $v^2$ method is impractical, and which approaches $\sigma$ in our far-from-equilibrium examples.

When analyzing experimental data, we recommend applying both methods and comparing their predictions. Even better, one should enlarge the pool of tested methods. Possible alternatives include saturating thermodynamic lower bounds with machine learning~\cite{manikandan2020inferring,otsubo2020estimating,manikandan2021quantitative,manikandan2024estimate,otsubo2022estimating,aguilera2026inferring,das2026localising,manikandan2026free}, or first inferring forces and diffusion parameters with stochastic force inference~\cite{frishman2020learning,bruckner2020inferring,hem2025learning} and then reconstructing $\sigma$ with Eq.~\eqref{sigma-force-discr}. Future studies should also compare CYNAR with these methods.

The CYNAR package is available as a GitHub repository~\cite{cynar_github}. Its code also supports analyzing displacement covariances in place of correlation functions. Displacement covariances were originally used to estimate entropy production via the variance sum rule~\cite{diterlizzi2024variance,diterlizzi2024vsrmodels}, from which the approach of Ref.~\cite{diterlizzi2025force} was derived. In forthcoming work, we will extend CYNAR to unbounded systems, which do not display well-defined steady-state position correlation functions but where displacement covariances remain meaningful. Other important directions include extending to state-dependent diffusion, for which the traffic can no longer be obtained from the short-time derivatives of the correlation functions, and estimating dissipation in many-body active-matter systems, a setting that will require additional information-theoretic approximations to infer the inflow rate.

\begin{acknowledgments}
M.B.~acknowledges hospitality and support from MPI-PKS Dresden, where the first part of this work was carried out.
We thank Felix Ritort, Matthias Fuchs, Gianmaria Falasco and Edgar Rold{\'a}n for useful discussions.

\textit{Declaration of AI usage}: Initially, ChatGPT, then Claude (Sonnet 5 and Opus 5.5) were used for
assistance with coding and text editing. The authors reviewed and verified all outputs.

\end{acknowledgments}

\appendix
\section{Derivation of the improved bound}
\label{app:bound}

We derive a lower bound on $\sigma$ that only uses a subset $\subs$ of observed degrees of freedom. We denote by $\hidd$ the complementary set of hidden ones, so that $\xx=(\xx_\subs,\xx_\hidd)$, and write the diffusion tensor in blocks $\MD_\subs\equiv\MD_{\subs\subs}$, $\MD_{\subs \hidd}$, and $\MD_\hidd\equiv\MD_{\hidd\hidd}$. We only assume a constant, positive-definite $\MD$ and natural boundary conditions. The starting point is the steady-state entropy production rate written in terms of the local mean velocity~\eqref{velocity},
\begin{align}
\label{app:sigma-v}
\sigma = \mean{\velo\cdot\MD^{-1}\velo} \, ,
\end{align}
with $\velo = \drift - \MD\score$ and $\score = \nabla\ln\rho$.

We first show that restricting Eq.~\eqref{app:sigma-v} to the observed components can only decrease it. For any two vectors $\bm u$ and $\bm y$, expanding the square shows that
\begin{equation}
\begin{split}
&2\,\bm y\cdot\bm u - \bm y\cdot\MD\bm y \\
&= \bm u\cdot\MD^{-1}\bm u - (\bm y-\MD^{-1}\bm u)\cdot\MD(\bm y-\MD^{-1}\bm u) \, .
\end{split}
\end{equation}
Since $\MD$ is positive definite, the last term is non-negative, and therefore
\begin{equation}
\label{app:var}
2\,\bm y\cdot\bm u - \bm y\cdot\MD\bm y \le \bm u\cdot\MD^{-1}\bm u
\end{equation}
for every $\bm y$, with equality for $\bm y=\MD^{-1}\bm u$. We now choose $\bm y=(\bm y_\subs,\bm 0)$. In this case, only the observed block of $\MD$ enters the left-hand side of Eq.~\eqref{app:var},
\begin{equation}
2\,\bm y\cdot\bm u - \bm y\cdot\MD\bm y = 2\,\bm y_\subs\cdot\bm u_\subs - \bm y_\subs\cdot\MD_\subs\bm y_\subs \, .
\end{equation}
The observed block $\MD_\subs$ is positive definite as well, since $\bm y_\subs\cdot\MD_\subs\bm y_\subs=(\bm y_\subs,\bm 0)\cdot\MD(\bm y_\subs,\bm 0)>0$ for any $\bm y_\subs\neq0$. Hence, $\MD_\subs^{-1}$ exists, and setting $\bm y_\subs=\MD_\subs^{-1}\bm u_\subs$, the right-hand side becomes $2\,\bm u_\subs\cdot\MD_\subs^{-1}\bm u_\subs-\bm u_\subs\cdot\MD_\subs^{-1}\bm u_\subs=\bm u_\subs\cdot\MD_\subs^{-1}\bm u_\subs$. Inserting this into Eq.~\eqref{app:var} gives
\begin{equation}
\bm u_\subs\cdot\MD_\subs^{-1}\bm u_\subs \le \bm u\cdot\MD^{-1}\bm u \, . \label{app:restrict}
\end{equation}
Applying this inequality to $\velo(\xx)$ at every point and averaging over the full density $\rho(\xx)$, we obtain
\begin{equation}
\label{app:sigmaS}
\sigma \;\ge\; \sigma^\subs \equiv \mean{\velo_\subs\cdot\MD_\subs^{-1}\velo_\subs} \, .
\end{equation}
Here, $\velo_\subs$ collects the observed components of the full velocity field, which depends on both $\xx_\subs$ and $\xx_\hidd$. Note that $\MD_\subs^{-1}$ is the inverse of the observed block of $\MD$, which differs from the observed block of $\MD^{-1}$ when $\MD_{\subs \hidd}\neq0$. This is the relevant quantity in practice, since $\MD_\subs$ follows from the correlations of $\xx_\subs$ alone through $\MD_\subs=-\dot\MC_{*,\subs}(0^+)$, and inverting it requires no knowledge of $\MD_{\subs \hidd}$ or $\MD_\hidd$, whereas the observed block of $\MD^{-1}$ would require the full diffusion tensor. When $\MD_{\subs \hidd}\neq0$, $\sigma$ does not split into contributions of $\subs$ and $\hidd$, but Eq.~\eqref{app:sigmaS} still holds.

The traffic of the observed degrees of freedom is defined, as in Eq.~\eqref{TS}, through the correlations of $\xx_\subs$,
\begin{align}
\label{app:TS}
\traffic^\subs = -\frac14\,\Tr\left[\MD_\subs^{-1}\,\ddot\MC_{*,\subs}(0^+)\right] ,
\end{align}
with $\MD_\subs = -\dot\MC_{*,\subs}(0^+)$. To relate it to $\sigma^\subs$, we express the short-time derivatives of the correlation functions through the drift. Since $\mean{x^i}\mean{x^j}$ does not depend on time in the steady state, only $\mean{x^i(t)x^j_0}$ contributes to the derivative terms. By It\^o's formula, $\frac{d}{dt}\mean{f(\xx_t)\,x^j_0}=\mean{(\mathcal{L}f)(\xx_t)\,x^j_0}$ for any smooth $f$, with $\mathcal{L}f=\drift\cdot\nabla f+\MD:\nabla\nabla f$. Choosing $f=x^i$ and then $f=a^i$, and taking $t\to0^+$, gives $\dot C^{ij}(0^+) = \mean{a^i x^j}$ and
\begin{align}
\ddot C^{ij}(0^+) = \mean{x^j\left(\drift\cdot\nabla a^i + \MD:\nabla\nabla a^i\right)} \, ,
\end{align}
where all averages are now taken at equal times over $\rho(\xx)$. We integrate the diffusion term by parts,
\begin{equation}
\begin{split}
\mean{x^j\,\MD:\nabla\nabla a^i}
&= -\int d\xx\;\partial_k(\rho\,x^j)\,D^{kl}\,\partial_l a^i \\
&= -\mean{x^j\,(\MD\score)\cdot\nabla a^i} - \mean{(\MD\nabla a^i)^j} \, ,
\end{split}
\end{equation}
where we used $\partial_k(\rho\,x^j)=\rho\,x^j s^k+\rho\,\delta^{jk}$ and summation over repeated indices. Combining it with the drift term and using $\velo=\drift-\MD\score$, we obtain
\begin{align}
\ddot C^{ij}(0^+) = \mean{x^j\,\velo\cdot\nabla a^i} - \mean{(\MD\nabla a^i)^j} \, .
\end{align}
Since $\rho\,\velo=\JJ$, the first term can be integrated by parts once more,
\begin{equation}
\begin{split}
\mean{x^j\,\velo\cdot\nabla a^i}
&= \int d\xx\; x^j\,\JJ\cdot\nabla a^i
= -\int d\xx\; a^i\,\nabla\cdot(x^j\JJ) \\
&= -\int d\xx\; a^i\left(J^j + x^j\,\nabla\cdot\JJ\right)
= -\mean{a^i v^j} \, ,
\end{split}
\end{equation}
where we used $\nabla\cdot\JJ=0$ in the steady state. Similarly, the second term gives
\begin{equation}
\begin{split}
-\mean{(\MD\nabla a^i)^j}
&= -\int d\xx\;\rho\,D^{jl}\,\partial_l a^i \\
&= \int d\xx\; a^i\,D^{jl}\,\partial_l\rho
= \mean{a^i\,(\MD\score)^j} \, .
\end{split}
\end{equation}
Writing $v^j=a^j-(\MD\score)^j$ in the first term and summing the two contributions, we find
\begin{align}
\ddot C^{ij}(0^+) = -\mean{a^i a^j} + 2\mean{a^i\,(\MD\score)^j} \, .
\end{align}
The same procedure applied to the stationarity condition $\frac{d}{dt}\mean{x^ix^j}=\mean{a^ix^j+a^jx^i}+2D^{ij}=0$ yields $\dot C_*^{ij}(0^+)=-D^{ij}$, i.e., Eq.~\eqref{Dhat}.

Only the observed block $\ddot\MC_{*,\subs}(0^+)$, i.e., the elements with $i,j\in \subs$, enters Eq.~\eqref{app:TS}. For these elements, only the observed components of the vector $\MD\score$ appear, which we collect in
\begin{align}
\bm w \equiv (\MD\score)_\subs = \MD_\subs\score_\subs + \MD_{\subs \hidd}\score_\hidd \, .
\end{align}
Each component $w^j=\sum_k D^{jk}s^k$ runs over all degrees of freedom, which is why the hidden score $\score_\hidd$ appears through the off-diagonal block $\MD_{\subs \hidd}$. In matrix form, the observed block of the symmetrized second derivative then reads
\begin{equation}
\begin{split}
\ddot\MC_{*,\subs}(0^+) = &-\mean{\drift_\subs\,\drift_\subs^T} + \mean{\drift_\subs\,\bm w^T} + \mean{\bm w\,\drift_\subs^T} \, .
\end{split}
\end{equation}
Inserting this expression into Eq.~\eqref{app:TS}, the two cross terms give equal contributions by the cyclicity of the trace, and we find
\begin{align}
4\traffic^\subs = \mean{\drift_\subs\cdot\MD_\subs^{-1}\drift_\subs} - 2\mean{\drift_\subs\cdot\MD_\subs^{-1}\bm w} \, .
\end{align}
Since $\velo_\subs = \drift_\subs - \bm w$, expanding Eq.~\eqref{app:sigmaS} leads to the exact decomposition
\begin{align}
\label{app:sigmaS-decomp}
\sigma^\subs = 4\traffic^\subs + \mean{\bm w\cdot\MD_\subs^{-1}\bm w} \, ,
\end{align}
which reduces to $\sigma=4\traffic+\inflow$ when all degrees of freedom are observed. For a proper subset $\subs$, however, $\bm w$ depends on the hidden coordinates through the full score, and the last term cannot be measured only from the trajectory of $\xx_\subs$.

The observed coordinates follow the marginal density $\rho_\subs(\xx_\subs)=\int d\xx_\hidd\,\rho(\xx)$, with marginal score $\tilde\score=\nabla_\subs\ln\rho_\subs$. Since $\nabla_\subs\rho_\subs=\int d\xx_\hidd\,\nabla_\subs\rho$, the marginal score is the conditional average of the full one, $\tilde\score(\xx_\subs) = \mean{\score_\subs\,|\,\xx_\subs}$. Moreover, natural boundary conditions give
\begin{align}
\mean{\score_\hidd\,|\,\xx_\subs}=\int d\xx_\hidd\,\frac{\rho}{\rho_\subs}\,\frac{\nabla_\hidd\rho}{\rho}=\frac{1}{\rho_\subs}\int d\xx_\hidd\,\nabla_\hidd\rho=0 \, ,
\end{align}
since the integral of a gradient over the hidden coordinates reduces to a vanishing boundary term. Because $\MD_\subs$ and $\MD_{\subs \hidd}$ are constant, the conditional average of $\bm w$ then reads
\begin{align}
\mean{\bm w\,|\,\xx_\subs} = \MD_\subs\,\mean{\score_\subs\,|\,\xx_\subs} + \MD_{\subs \hidd}\,\mean{\score_\hidd\,|\,\xx_\subs} = \MD_\subs\,\tilde\score(\xx_\subs) \, .
\end{align}
We write $\bm w=\MD_\subs\tilde\score+\bm{\delta w}$, where the fluctuation $\bm{\delta w}=\MD_\subs(\score_\subs-\tilde\score)+\MD_{\subs \hidd}\score_\hidd$ has vanishing conditional average, $\mean{\bm{\delta w}\,|\,\xx_\subs}=0$. Since $\tilde\score$ only depends on $\xx_\subs$, the cross term $\mean{\tilde\score\cdot\bm{\delta w}}=\mean{\tilde\score\cdot\mean{\bm{\delta w}\,|\,\xx_\subs}}$ vanishes, and
\begin{align}
\mean{\bm w\cdot\MD_\subs^{-1}\bm w}
= \inflow^\subs + \mean{\bm{\delta w}\cdot\MD_\subs^{-1}\bm{\delta w}}
\;\ge\; \inflow^\subs \, ,
\end{align}
where we defined the inflow rate of the observed degrees of freedom,
\begin{align}
\label{app:GS}
\inflow^\subs \equiv \mean{\tilde\score\cdot\MD_\subs\tilde\score}
= \int d\xx_\subs\,\rho_\subs\,\tilde\score\cdot\MD_\subs\tilde\score \, .
\end{align}
The coupling $\MD_{\subs \hidd}$ thus only contributes to the non-negative fluctuation term, which is discarded in the bound. Since $\mean{\score\score^T}$ and $\mean{\tilde\score\tilde\score^T}$ are Fisher information matrices with respect to translations of the state, this inequality expresses the loss of Fisher information upon marginalization over the hidden degrees of freedom. Moreover, applying Eq.~\eqref{app:restrict} to $\bm u=\MD\score$, whose observed components are $\bm w$, gives $\mean{\bm w\cdot\MD_\subs^{-1}\bm w}\le\mean{\score\cdot\MD\score}=\inflow$. Hence, the gap between $\sigma^\subs$ and the bound satisfies
\begin{align}
\label{app:gap}
0\;\le\;\sigma^\subs-\left(4\traffic^\subs+\inflow^\subs\right)=\mean{\bm{\delta w}\cdot\MD_\subs^{-1}\bm{\delta w}}\;\le\;\inflow-\inflow^\subs \, ,
\end{align}
i.e., it cannot exceed the part of the inflow rate lost upon marginalization over the hidden degrees of freedom. More precisely, expanding $\bm w=\MD_\subs\score_\subs+\MD_{\subs\hidd}\score_\hidd$ shows that the last inequality falls short by $\mean{\score_\hidd\cdot(\MD_\hidd-\MD_{\hidd\subs}\MD_\subs^{-1}\MD_{\subs\hidd})\,\score_\hidd}>0$, the hidden block of the Fisher information weighted by the positive-definite Schur complement of $\MD_\subs$.

Combining Eqs.~\eqref{app:sigmaS}, \eqref{app:sigmaS-decomp}, and \eqref{app:GS} with $\sigma\ge0$, we obtain the improved bound
\begin{align}
\label{app:bound-final}
\sigma \;\ge\; \sigma^\subs \;\ge\; \max\left(4\traffic^\subs+\inflow^\subs,\,0\right) .
\end{align}
Both terms follow from the observed trajectory: the traffic~\eqref{app:TS} from the correlations of $\xx_\subs$, and the inflow rate~\eqref{app:GS} from their histogram. In particular, no knowledge of $\MD_{\subs \hidd}$ or $\MD_\hidd$ is required, since $\MD_\subs$ is obtained from the observed correlations. The bound holds for any positive-definite $\MD$, becomes an equality when all degrees of freedom are observed and, since $\inflow^\subs\ge0$, it is tighter than $\max(4\traffic^\subs,0)$. The same argument applies separately to each block of a partition of $\subs$ that $\MD_\subs$ does not couple. For a diagonal $\MD_\subs$, in particular,
\begin{align}
\label{app:bound-diag}
\sigma \;\ge\; \sum_{i\in \subs}\max\left(4\traffic^i+\inflow^i,\,0\right) ,
\end{align}
with $\traffic^i = -\ddot C_*^{ii}(0^+)/(4D^{ii})$ and $\inflow^i = D^{ii}\mean{(\partial_i\ln\rho_i)^2}$, where $\rho_i$ is the marginal density of $x^i$. This improves the bound $\sigma\ge\sum_{i\in \subs}\max(4\traffic^i,0)$ based on only the traffic.

\section{The $v^2$ method under partial observation}
\label{app:v2}

We now consider the $v^2$ method applied to the observed coordinates $\xx_\subs$, using the notation of Appendix~\ref{app:bound}. Integrating the Fokker-Planck equation~\eqref{FP} over $\xx_\hidd$, the term $\nabla_\hidd\cdot\JJ_\hidd$ reduces to a vanishing boundary term, so that the marginal density obeys a continuity equation with current $\tilde\JJ_\subs=\int d\xx_\hidd\,\JJ_\subs$. The corresponding velocity,
\begin{align}
\tilde\velo=\frac{\tilde\JJ_\subs}{\rho_\subs}=\int d\xx_\hidd\,\frac{\rho}{\rho_\subs}\,\velo_\subs=\mean{\velo_\subs\,|\,\xx_\subs} \, ,
\end{align}
is the quantity estimated by binning the observed displacements, Eq.~\eqref{v-binned}. Since $\MD_\subs^{-1}$ is constant and $\mean{\velo_\subs-\tilde\velo\,|\,\xx_\subs}=0$, the cross term vanishes in
\begin{equation}
  \begin{split}
\mean{\velo_\subs\cdot\MD_\subs^{-1}\velo_\subs\,|\,\xx_\subs}
=&\tilde\velo\cdot\MD_\subs^{-1}\tilde\velo
  \\
  &+\mean{(\velo_\subs-\tilde\velo)\cdot\MD_\subs^{-1}(\velo_\subs-\tilde\velo)\,|\,\xx_\subs} \, ,
  \end{split}
\end{equation}
and averaging over $\xx_\subs$ gives $\sigma_{v^2}^\subs\le\sigma^\subs$, Eq.~\eqref{sigmav2S}. The gap is the $\MD_\subs^{-1}$-weighted conditional variance of $\velo_\subs$, which vanishes only if the observed velocity does not depend on the hidden coordinates. For a single observed coordinate, stationarity makes the marginal current constant, and natural boundary conditions set it to zero, so that $\tilde\velo=0$, whereas $4\traffic^\subs+\inflow^\subs$ can remain positive.

\bibliography{biblio_paper1}

\end{document}